\documentclass[12pt,showpacs,preprintnumbers,amsmath,amssymb,eqsecnum]{revtex4}
\usepackage{graphicx}
\begin{document}

\newcommand{\pderiv}[2]{\frac{\partial #1}{\partial #2}}
\newcommand{\deriv}[2]{\frac{d #1}{d #2}}

\title{Multicritical Behavior in a Random-Field Ising Model under a
Continuous-Field Probability Distribution}

\vskip \baselineskip

\author{Octavio R. Salmon$^{1}$}
\thanks{E-mail: octavior@cbpf.br}

\author{Nuno Crokidakis$^{2}$}
\thanks{E-mail: nuno@if.uff.br}

\author{Fernando D. Nobre$^{1}$}
\thanks{Corresponding author: fdnobre@cbpf.br}

\address{
$^{1}$Centro Brasileiro de Pesquisas F\'{\i}sicas \\
Rua Xavier Sigaud 150 \\
22290-180 \hspace{5mm} Rio de Janeiro - RJ \hspace{5mm} Brazil \\
$^{2}$Instituto de F\'{\i}sica - Universidade Federal Fluminense \\
Av. Litor\^anea, s/n \\
24210-340 \hspace{5mm} Niter\'oi - RJ \hspace{5mm} Brazil}

\date{\today}


\begin{abstract}
\noindent
A random-field Ising model that is capable of exhibiting a rich variety of
multicritical phenomena, as well as a smearing of such behavior, is investigated. The model consists of an
infinite-range-interaction Ising ferromagnet in the presence of a 
triple-Gaussian random magnetic field, which is defined as a superposition of
three Gaussian distributions with the same width $\sigma$, 
centered at $H=0$ and $H=\pm H_{0}$, with
probabilities $p$ and $(1-p)/2$, respectively. Such a distribution is very
general and recovers 
as limiting cases, the trimodal, bimodal, and Gaussian probability 
distributions. In
particular, the special case of the random-field Ising model in the 
presence of a trimodal
probability distribution (limit $\sigma \rightarrow 0$) is able to present
a rather nontrivial multicritical behavior. It is argued that the
triple-Gaussian probability 
distribution is appropriate for a physical description of some diluted
antiferromagnets in the presence of a uniform external field, for which the
corresponding physical realization consists of an Ising ferromagnet under
random fields whose distribution appears to be well-represented in terms of a
superposition of two parts, namely, a trimodal and a continuous contribution.  
The model is investigated
by means of the replica method, and phase diagrams are obtained within the
replica-symmetric solution, which is known to be stable for the present
system.
A rich variety of phase diagrams is presented, with a single or
two distinct ferromagnetic phases, continuous and first-order 
transition lines, tricritical, fourth-order, critical end
points, and many other interesting multicritical phenomena.
Additionally, the present model carries the
possibility of destructing such multicritical phenomena due to an increase 
in the randomness, i.e., increasing $\sigma$, which represents a very 
common feature in real systems. 

\vskip \baselineskip


\end{abstract}
\maketitle

\vskip \baselineskip

\section{INTRODUCTION}

The Random Field Ising Model (RFIM) 
became nowadays one of the most studied problems in the area of
disordered magnetic  
systems \cite{youngbook,dotsenkobook}. From the theoretical point of view 
\cite{nattermannreview}, its simple definition \cite{imryma}, together with the richness of physical properties that emerge from its 
study, represent two main motivations
for the investigation of this model. On the other hand, a
considerable experimental interest \cite{belangerreview} arised after 
the identification of the RFIM with diluted antiferromagnets
in the presence of a uniform magnetic field 
\cite{fishmanaharony,pozenwong,cardy}; since then, two of the most
investigated systems are the compounds  
${\rm Fe_{x}Zn_{1-x}F_{2}}$ and
${\rm Fe_{x}Mg_{1-x}Cl_{2}}$ \cite{belangerreview,birgeneau}. 

In what concerns the equilibrium phase diagrams of the RFIM, 
the effects of different probability-distribution functions (PDFs)
for the random fields 
have attracted the attention of many authors, see, e.g., Refs.  
\cite{schneiderpytte,aharony78,andelman,galambirman,mattis,kaufman,%
nuno08a,aizenman,huiberker,gofman,swift,machta00,middleton,machta03,%
wumachta05,wumachta06,fytas}, among others. At the mean-field level,
the Gaussian PDF yields a continuous
ferromagnetic-paramagnetic boundary \cite{schneiderpytte}, whereas discrete
PDFs may lead to elaborate phase diagrams,
characterized by a finite-temperature tricritical point followed by a 
first-order phase transition at low temperatures 
\cite{aharony78,andelman,galambirman,mattis,kaufman}, 
or even fourth-order and critical end points \cite{mattis,kaufman}. For
the short-range-interaction RFIM, the existence of a 
low-temperature first-order phase transition remains a very controversial
issue \cite{aizenman,huiberker,gofman,swift,machta00,middleton,machta03,%
wumachta05,wumachta06,fytas}. 

From the experimental point of view, one is certainly concerned with 
physical realizations of the RFIM, for which the most commonly known are 
diluted antiferromagnets
in the presence of a uniform magnetic field \cite{belangerreview}. 
Hence, in the identification of RFIMs with such real systems
\cite{fishmanaharony,pozenwong,cardy}, certain classes of 
distributions for the random fields, in the corresponding RFIM, 
are more appropriate for a description of diluted 
antiferromagnets in the presence of a uniform magnetic field.
In the later systems one has local
variations in the sum of exchange couplings that connect a given site to
other sites, leading to local variations of the
two-sublattice site magnetizations, and as a consequence, one may have local
magnetizations that vary in both sign and magnitude. 
In the identifications of the RFIM with diluted antiferromagnets
\cite{fishmanaharony,pozenwong,cardy}, the effective random field at a 
given site is expressed always in terms of quantities that vary in both
sign and magnitude: 
(i) the local magnetization \cite{fishmanaharony};
(ii) two contributions, namely, a first one that
assumes only three discrete values, related to
the dilution of the system and the uniform external field, and a second one
that is proportional to the local magnetization \cite{pozenwong}; 
(iii) the sum of the exchange couplings associated to this 
site \cite{cardy}. Therefore, for a proper description of diluted
antiferromagnets in the presence of a uniform field, the corresponding 
RFIMs should always be considered in terms of continuous PDFs for the fields. 

The compound ${\rm Fe_{x}Mg_{1-x}Cl_{2}}$ presents an 
Ising spin-glass behavior for ${\rm x} < 0.55$, and is considered as a
typical RFIM 
for higher magnetic concentrations. In the RFIM regime, it shows 
some curious behavior and is considered as a candidate for exhibiting 
multicritical phenomena \cite{belangerreview,pozenwong,kaufman}.
As an example of this type of effect, one finds a
first-order transition turning into a continuous one 
due to a change in the random fields
\cite{belangerreview,kushauerkleemann,kushauer}; the concentration at which
the first-order transition disappears is estimated to be $x=0.6$. 
The crossover from first-order to continuous phase
transitions has been investigated through different
theoretical approaches \cite{aizenman,huiberker,kushauer}. One
possible mechanism used to find such a crossover, or even to suppress the 
first-order
transition completely, consists in introducing an additional kind of 
randomness in the system, e.g., bond randomness \cite{aizenman,huiberker}. 
By considering randomness in the field only, this crossover
has been also analyzed 
through zero-temperature studies, either within mean-field theory
\cite{sethna93}, or numerical simulations on a three-dimensional lattice
\cite{kushauer,sethna93}. Recently a RFIM has been proposed \cite{nuno08a}
for which the  
finite-temperature tricritical point, together with the first-order line, 
may disappear due to an 
increase in the field randomness, similarly to what happens with the
first-order phase transition in the compound 
${\rm Fe_{x}Mg_{1-x}Cl_{2}}$.

However,  it is possible that the diluted antiferromagnet 
${\rm Fe_{x}Mg_{1-x}Cl_{2}}$, or some other similar compound, 
may present an even more 
complicated critical behavior, not yet verified experimentally, to our knowledge. According to 
the analysis of Ref.~\cite{pozenwong}, one of its
contributions for the random fields assumes only the values 
$0, \pm \sqrt{2}H$ ($H$ represents the external uniform
magnetic field); motivated by this result, a RFIM was proposed 
\cite{mattis,kaufman} with the
random fields described in terms of a trimodal distribution. 
Such a RFIM, studied within a mean-field approach through a model defined 
in the limit of infinite-range interactions, yielded 
a rich critical behavior with the
occurence of first-order phase transitions, tricritical and 
high-order critical points \cite{griffiths} at finite
temperatures, and even the possibility of two distinct 
ferromagnetic phases at low temperatures \cite{mattis,kaufman}. 
These investigations suggest that ${\rm Fe_{x}Mg_{1-x}Cl_{2}}$ may 
exhibit a rich critical behavior that goes beyond 
the disappearance of a first-order phase transition observed around 
$x=0.6$.  
However, taking into account the above
criteria for the identification of the RFIM with diluted antiferromagnets, 
one notices that the trimodal distribution does not represent an
appropriate choice from the physical point of view. In fact, in the 
analysis of Ref.~\cite{pozenwong}, a second, \underline{continuous} 
contribution for the random fields should be taken into account as well.
Therefore, for  
an adequate theoretical description of ${\rm Fe_{x}Mg_{1-x}Cl_{2}}$, 
one should consider a RFIM defined in terms of a finite-width three-peaked
distribution, which could exhibit multicritical behavior,
and in addition to that, a crossover from such behavior to continuous phase
transitions due to an increase in the field randomness. 

For that purpose, herein we introduce a RFIM that  
consists of an
infinite-range-interaction Ising ferromagnet in the presence of a 
triple-Gaussian random magnetic field. This distribution is defined as a 
superposition of three Gaussian distributions with the same width $\sigma$, 
centered at $H=0$ and $H=\pm H_{0}$, with
probabilities $p$ and $(1-p)/2$, respectively. The particular case of the RFIM
in the presence of a trimodal PDF, studied previously 
\cite{mattis,kaufman}, is recovered in the limit $\sigma \rightarrow 0$. 
We show that the present model is capable of exhibiting a 
rich multicritical behavior, with phase diagrams displaying one, or even two
distinct ferromagnetic phases, 
continuous and first-order critical frontiers, tricritical points and
high-order critical points, at both finite and zero temperatures.
Such a variety of phase diagrams may be obtained from this model by varying the parameters of the corresponding PDF, e.g., $p$ and $\sigma$: 
(i) Variations in $p$ ($\sigma$ fixed) lead to qualitatively distinct multicritical phenomena; (ii) Increasing $\sigma$ ($p$ fixed) yields a smearing of an 
specific type of critical behavior, destroying a possible multicritical phenomena due to an increase in the randomness, representing a very 
common feature in real systems. 
In the next section we define the model, find its free energy and equation for 
the magnetization, by using the replica approach. In section III we present
several phase diagrams, characterized by the critical behavior mentioned
above. Finally, in section IV we present our conclusions.  

\section{THE MODEL}

The infinite-range-interaction Ising model in the presence of an external
random magnetic field is defined in terms of the Hamiltonian 

\begin{equation} \label{1}
\mathcal{H}=- \frac{J}{N}\sum_{(i,j)}S_{i}S_{j} - \sum_{i}H_{i}S_{i}~, 
\end{equation}

\vskip \baselineskip
\noindent
where the sum $\sum_{(i,j)}$ runs over all distinct pairs of spins
$S_{i}=\pm 1$ ($i=1,2,...,N$). The random fields $\{H_{i}\}$ are quenched
variables and obey the PDF, 

\begin{eqnarray}\nonumber 
P(H_{i}) & = & \frac{(1-p)}{2}\left(\frac{1}{2\pi \sigma^{2}}\right)^{1/2}
\left\{\exp\left[-\frac{(H_{i}-H_{0})^{2}}{2\sigma^{2}}\right]
+\exp\left[-\frac{(H_{i}+H_{0})^{2}}{2\sigma^{2}}\right]\right\} \\ \label{2}
& + & p\left(\frac{1}{2\pi \sigma^{2}}\right)^{1/2}
\exp\left[-\frac{H_{i}^{2}}{2\sigma^{2}}\right]~, 
\end{eqnarray}

\vskip \baselineskip
\noindent
which consists of a superposition of three independent Gaussian
distributions with the same width $\sigma$, 
centered at $H_{i}=0$ and $H_{i}=\pm H_{0}$, with
probabilities $p$ and $(1-p)/2$, respectively (to be called hereafter,
a triple Gaussian PDF).
This distribution is very general, depending on three parameters, namely, 
$p$, $\sigma$, and $H_{0}$, and contains as particular cases several
well-known distributions of the literature, namely the trimodal
and bimodal distributions, the double Gaussian, as well as simple Gaussian
distributions. The model defined above is expected to be appropriate for a
physical description of diluted antiferromagnets that in the presence of an 
external magnetic field may exhibit multicritical behavior; one candidate
for this purpose is ${\rm Fe_{x}Mg_{1-x}Cl_{2}}$ \cite{pozenwong}.  

From the free energy $F(\{H_{i}\})$, 
associated with a given realization of 
site fields $\{H_{i}\}$, one may calculate the quenched average, 
$[F(\{H_{i}\})]_{H}$, 

\begin{equation}  \label{3}
[F(\{H_{i}\})]_{H}=\int\prod_{i}[dH_{i}P(H_{i})]F(\{H_{i}\})~. 
\end{equation}

\vskip \baselineskip
\noindent
The standard procedure for carrying the average above is by making use of the 
replica method
\cite{dotsenkobook,nishimoribook}, leading to the free energy per spin,  

\begin{equation} \label{4}
-\beta f=\lim_{N \to \infty}\frac{1}{N}[\ln Z(\{H_{i}\})]_{H} 
= \lim_{N \to \infty}\lim_{n \to 0}\frac{1}{Nn}([Z^{n}]_{H}-1)~,
\end{equation}

\vskip \baselineskip
\noindent
where $Z^{n}$ is the partition function of $n$ copies of the original
system defined in Eq.~(\ref{1}) and $\beta = 1/(kT)$. One gets that, 

\begin{equation}  \label{5}
\beta f= \lim_{n \to 0} \frac{1}{n} \, {\rm min} \; g(m^{\alpha})~, 
\end{equation}

\vskip \baselineskip
\noindent
with

\begin{eqnarray}  \nonumber
g(m^{\alpha}) & = & \frac{\beta J}{2}\sum_{\alpha}(m^{\alpha})^{2}
-\frac{(1-p)}{2}\ln{\rm Tr}_{\alpha}\exp(\mathcal{H}_{{\rm eff}}^{+}) 
-\frac{(1-p)}{2}\ln{\rm Tr}_{\alpha}\exp(\mathcal{H}_{{\rm eff}}^{-}) 
\\ \label{6} 
& - & p\ln{\rm Tr}_{\alpha}\exp(\mathcal{H}_{{\rm eff}}^{(0)})~, \\ \label{7}
\mathcal{H}_{{\rm eff}}^{\pm} & = & \beta J\sum_{\alpha}m^{\alpha}S^{\alpha} 
+ \beta \sigma \left( \sum_{\alpha}S^{\alpha} \right)^{2}
\pm \beta H_{0}\sum_{\alpha}S^{\alpha}~, \\ \label{8}
\mathcal{H}_{{\rm eff}}^{(0)} & = & \beta J\sum_{\alpha}m^{\alpha}S^{\alpha} 
+ \beta \sigma \left( \sum_{\alpha}S^{\alpha} \right)^{2}~.
\end{eqnarray}

\vskip \baselineskip
\noindent
In the equations above, $\alpha$ represents a replica label
($\alpha=1,2,...,n$) and ${\rm Tr}_{\alpha}$ stands for a trace over 
the spin variables of each replica.
The extrema of the functional $g(m^{\alpha})$ leads to the 
equation for the magnetization of replica $\alpha$, 

\begin{equation} \label{81}
m^{\alpha}=\frac{(1-p)}{2}<S^{\alpha}>_{+} \
+ \ \frac{(1-p)}{2}<S^{\alpha}>_{-} \ + \ p<S^{\alpha}>_{0}~, 
\end{equation}

\vskip \baselineskip
\noindent
where $<\,>_{\pm}$ and $<\,>_{0}$ denote thermal averages with respect to
the ``effective Hamiltonians'' $\mathcal{H}_{{\rm eff}}^{\pm}$ and
$\mathcal{H}_{{\rm eff}}^{(0)}$, in Eqs.~(\ref{7}) and (\ref{8}),
respectively.  

If one assumes the replica-symmetry ansatz
\cite{dotsenkobook,nishimoribook}, i.e., $m^{\alpha}=m$ ($\forall~\alpha$),
the free energy per spin of Eqs.~(\ref{5})--(\ref{8}) and the
equilibrium condition, Eq.~(\ref{81}), become 

\begin{eqnarray} \nonumber
f & = & \frac{J}{2}m^{2} - \frac{(1-p)}{2\beta}
\int Dz\ln(2\cosh\Phi^{+}) - \frac{(1-p)}{2\beta}\int Dz\ln(2\cosh\Phi^{-}) 
\\ \label{9}
& - & \frac{p}{2}\int Dz\ln(2\cosh\Phi^{(0)})~,  \\ \nonumber 
\\ \label{10}
m & = & \frac{(1-p)}{2}\int Dz\tanh\Phi^{+} + \frac{(1-p)}{2}
\int Dz\tanh\Phi^{-} +p\int Dz\tanh\Phi^{(0)}~,
\end{eqnarray}

\vskip \baselineskip
\noindent
where

\begin{equation}
\Phi^{\pm}=\beta (Jm + \sigma z \pm H_{0})~; \quad 
\Phi^{(0)}=\beta (Jm + \sigma z)~; \quad 
\int Dz\;(..) =  \frac{1}{\sqrt{2\pi}} 
\int_{-\infty}^{+\infty}dz\;e^{-z^{2}/2}\;(..)~.
\end{equation}

\noindent
It should be mentioned that the 
instability associated with the replica-symmetric solution \cite{at}
at low temperatures is usually related to parameters characterized by two
replica indices, 
like in the spin-glass problem \cite{dotsenkobook,nishimoribook}. In the
present system the order parameter $m^{\alpha}$ depends on a single replica
index, and so such an instability does not occur. 

In the next section we will make use of 
Eqs.~(\ref{9}) and (\ref{10}) in order to get several phase diagrams for
the present model. Although most critical frontiers will be 
achieved numerically, some analytical results may be obtained at zero
temperature.   
At $T=0$, the free energy and magnetization become,
respectively, 

\begin{eqnarray}\nonumber
f&=&-\frac{J}{2}m^{2}-\frac{H_{0}}{2}(1-p)
\left[{\rm erf}\left(\frac{Jm+H_{0}}{\sigma\sqrt{2}}\right) 
- {\rm erf}\left(\frac{Jm-H_{0}}{\sigma\sqrt{2}}\right)\right] \\ \nonumber
&-& \frac{\sigma}{\sqrt{2\pi}}(1-p)
\left\{\exp\left[-\frac{(Jm+H_{0})^{2}}{2\sigma^{2}}\right] +
\exp\left[-\frac{(Jm-H_{0})^{2}}{2\sigma^{2}}\right]\right\} \\
&-& \frac{2\sigma}{\sqrt{2\pi}}\;p
\exp\left[-\frac{(Jm)^{2}}{2\sigma^{2}}\right]~, 
\label{16}
\end{eqnarray}

\begin{equation}
m=\frac{(1-p)}{2}~{\rm erf}\left(\frac{Jm+H_{0}}{\sigma \sqrt{2}}\right) +
\frac{(1-p)}{2}~{\rm erf}\left(\frac{Jm-H_{0}}{\sigma \sqrt{2}}\right)+ 
p\;{\rm erf}\left(\frac{Jm}{\sigma \sqrt{2}}\right)~.
\label{17}
\end{equation}

\section{PHASE DIAGRAMS}

Usually, in the RFIM one has two phases, namely, the 
Ferromagnetic ({\bf F})
and Paramagnetic ({\bf P}) ones. However, for the PDF of 
Eq.~(\ref{2}), there is a possibility of three distinct
phases: apart from the Paramagnetic ($m = 0$), one has  
two Ferromagnetic phases, {\bf F}$\!_1$ and {\bf F}$\!_2$~, 
characterized by different magnetizations (herein we will consider always  
{\bf F}$\!_1$ as the ordered phase with higher magnetization, i.e., 
$m_{1} > m_{2} > 0$). 
Close to a
continuous transition between an ordered and the disordered phases, 
$m$ is small,
so that one can expand Eq.~(\ref{10}) in powers of $m$, 

\begin{equation} \label{11}
m = A_{1}m + A_{3}m^{3} + A_{5}m^{5} + A_{7}m^{7} + 0(m^{9})~, 
\end{equation}

\vskip \baselineskip
\noindent
where the coefficients are given by

\begin{eqnarray} \label{12}
A_{1} &=& \beta J\{1-(1-p)\lambda_{1}-p\lambda_{1}^{(0)}\}~, \\  \label{13}
A_{3} &=& -\frac{(\beta J)^{3}}{3}\{1-4[(1-p)\lambda_{1}+p\lambda_{1}^{(0)}]
+3[(1-p)\lambda_{2}+p\lambda_{2}^{(0)}]\}~,  \\ \nonumber
A_{5} &=& \frac{(\beta J)^{5}}{15}\{2-17[(1-p)\lambda_{1}+p\lambda_{1}^{(0)}]
+30[(1-p)\lambda_{2}+p\lambda_{2}^{(0)}]-15[(1-p)\lambda_{3} \\ \label{14}
&+& p\lambda_{3}^{(0)}]\}~ \\ \nonumber
A_{7} &=& \frac{(\beta J)^{7}}{315}\{17-248[(1-p)\lambda_{1}+p\lambda_{1}^{(0)}]
+756[(1-p)\lambda_{2}+p\lambda_{2}^{(0)}]-840[(1-p)\lambda_{3} \\
& + & p\lambda_{3}^{(0)}]+315[(1-p)\lambda_{4}+p\lambda_{4}^{(0)}]\}~,
\end{eqnarray}

\vskip \baselineskip
\noindent
with

\begin{eqnarray} \nonumber
\lambda_{k} & = & \frac{1}{\sqrt{2\pi}}\int_{-\infty}^{+\infty}dz\;e^{-z^{2}/2}
\tanh^{2k}\beta(H_{0}+\sigma z)~,\\ \label{15}
\lambda_{k}^{(0)} & = & \frac{1}{\sqrt{2\pi}} 
\int_{-\infty}^{+\infty}dz\;e^{-z^{2}/2}
\tanh^{2k}(\beta\sigma z)~.
\end{eqnarray}

\vskip \baselineskip
\noindent
In order to find the continuous critical frontier one sets $A_{1}=1$,
provided that $A_{3}<0$. If a first-order critical frontier also occurs, the
continuous line ends when $A_{3}=0$; in such cases, the continuous and
first-order critical frontiers meet at a tricritical point, whose
coordinates may be obtained by solving the equations $A_{1}=1$ and 
$A_{3}=0$
numerically, provided that $A_{5}<0$. 
In the present problem there is also a possibility of a 
fourth-order critical point, which is obtained from the conditions
$A_{1}=1$, $A_{3}=A_{5}=0$ and $A_{7}<0$. 
In addition, when the two distinct ferromagnetic phases are present, 
two other critical points may also appear (herein we
follow the classification due to Griffithis \cite{griffiths}): 
(i) the ordered critical point, which corresponds to an isolated critical
point inside the ordered region, terminating a first-order line that
separates phases {\bf F}$\!_1$ and {\bf F}$\!_2$~;
(ii) the critical end point, where all three phases coexist, corresponding
to the intersection of a continuous line that separates the paramagnetic
from one of the ferromagnetic phases with a first-order line separating the
paramagnetic and the other ferromagnetic phase.  
The location of the critical points defined in (i) and (ii), as well as 
the first-order critical frontiers, were determined by a numerical analysis
of the free-energy minima, e.g.,  
two equal minima for the free energy, characterized by 
two different values of
magnetization, $m_{1}>m_{2}>0$, yields a 
point of the first-order critical frontier separating phases  
{\bf F}$\!_1$ and {\bf F}$\!_2$~.

Next, we show several phase diagrams of this model
for both finite and zero temperatures.
Besides the notation defined above for labeling the phases,
in these phase diagrams we shall use distinct symbols and representations
for the critical points and frontiers, as described below.   

\begin{itemize}

\item Continuous critical frontier: continuous line;

\item First-order critical frontier: dotted line;

\item Tricritical point: located by a black circle;

\item Fourth-order point: located by an empty square;

\item Ordered critical point: located by a black asterisk;

\item Critical end point: located by a black triangle.

\end{itemize}

\subsection{Finite-Temperature Critical Frontiers}

We now present finite-temperature phase diagrams; these phase diagrams will
be exhibited in the plane of dimensionless variables, $(kT)/J$ versus 
$H_{0}/J$, for typical values of $p$ and $\sigma/J$. For completeness, in
each figure we also show the forms of the corresponding PDF 
of Eq.~(\ref{2}), with the parameters used in the phase diagrams. 

\begin{figure}[t]
\begin{center}
\includegraphics[width=0.35\textwidth,angle=0]{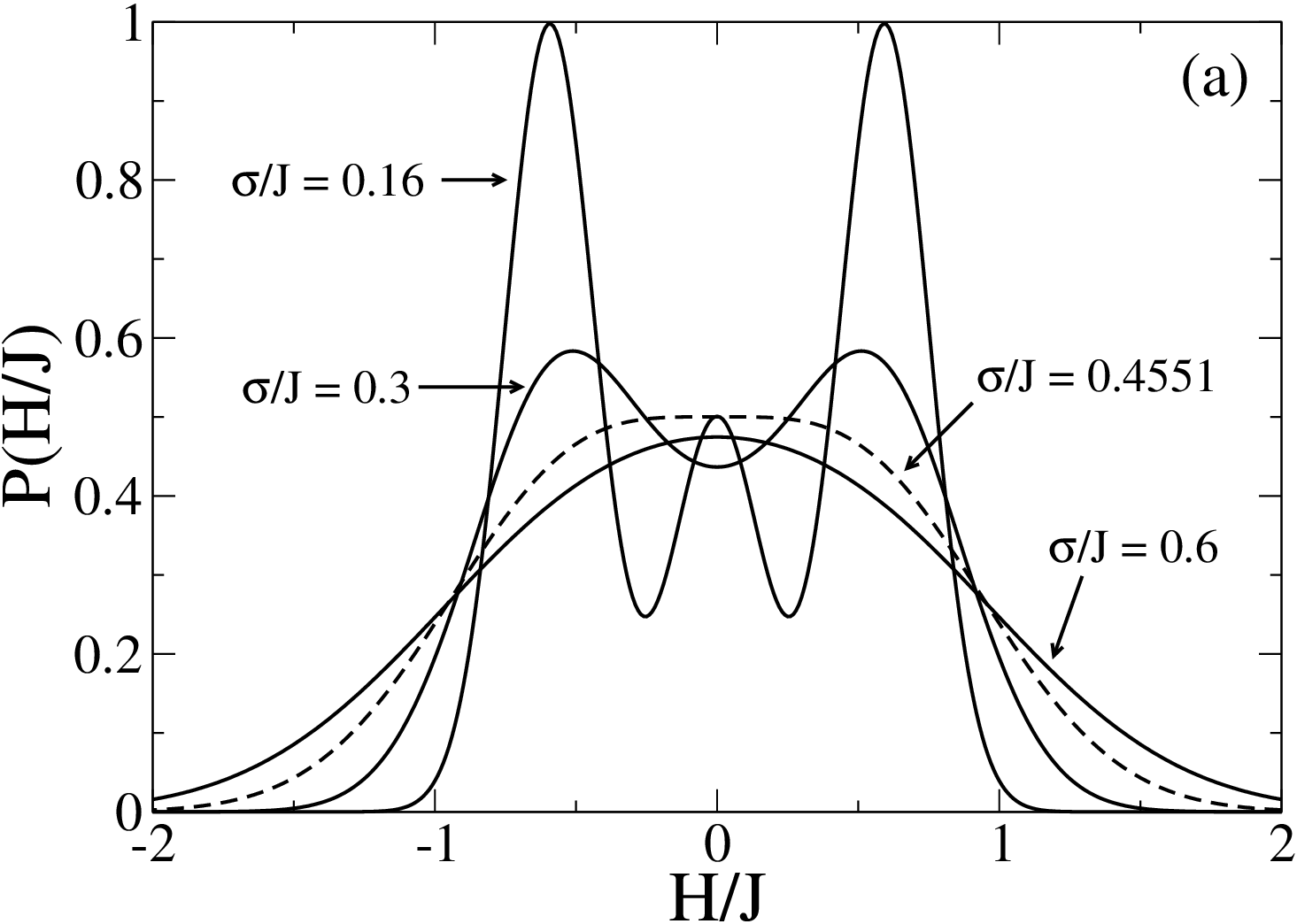}
\hspace{1.5cm}
\includegraphics[width=0.35\textwidth,angle=0]{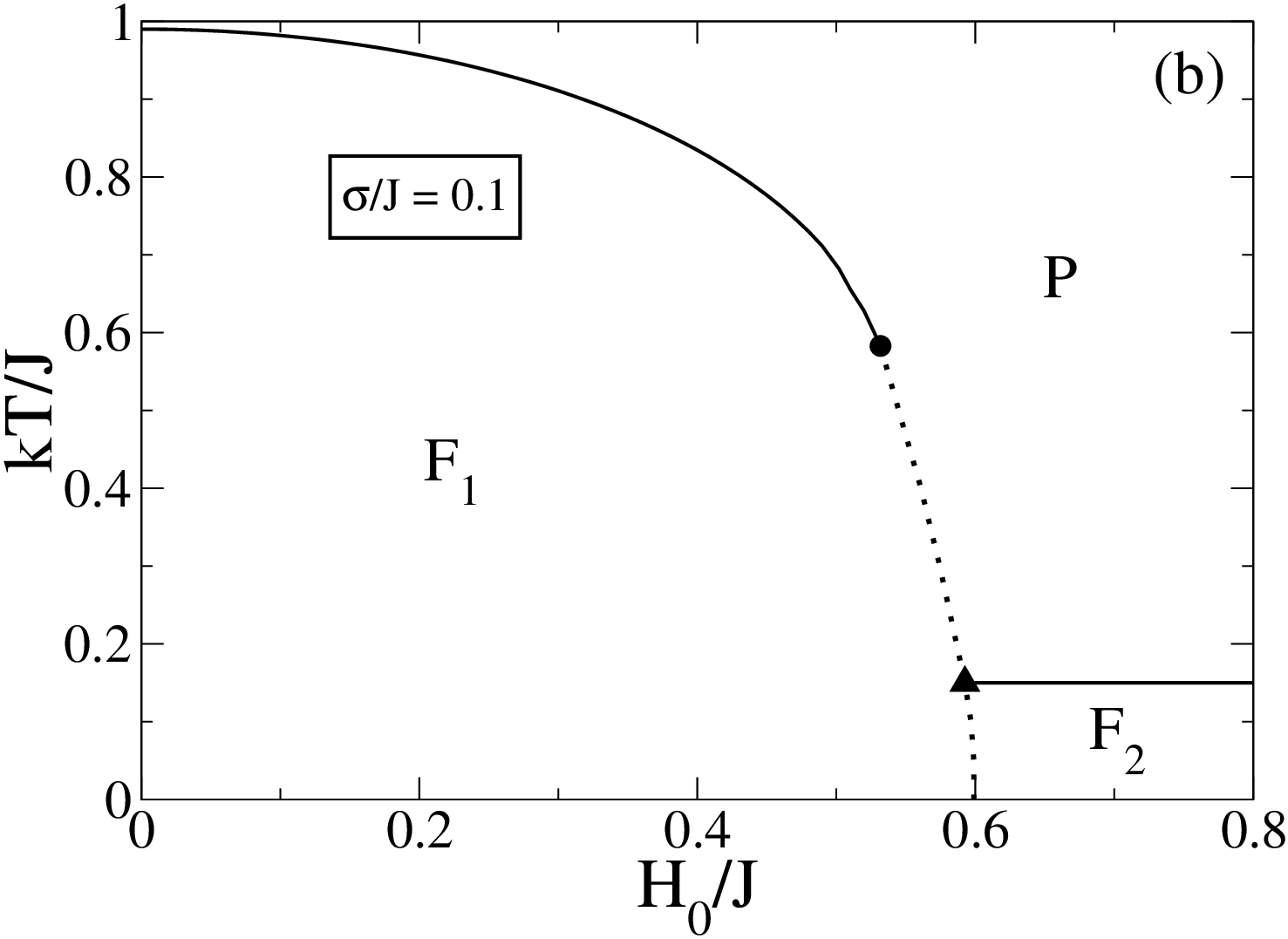}
\\
\vspace{1.0cm}
\includegraphics[width=0.35\textwidth,angle=0]{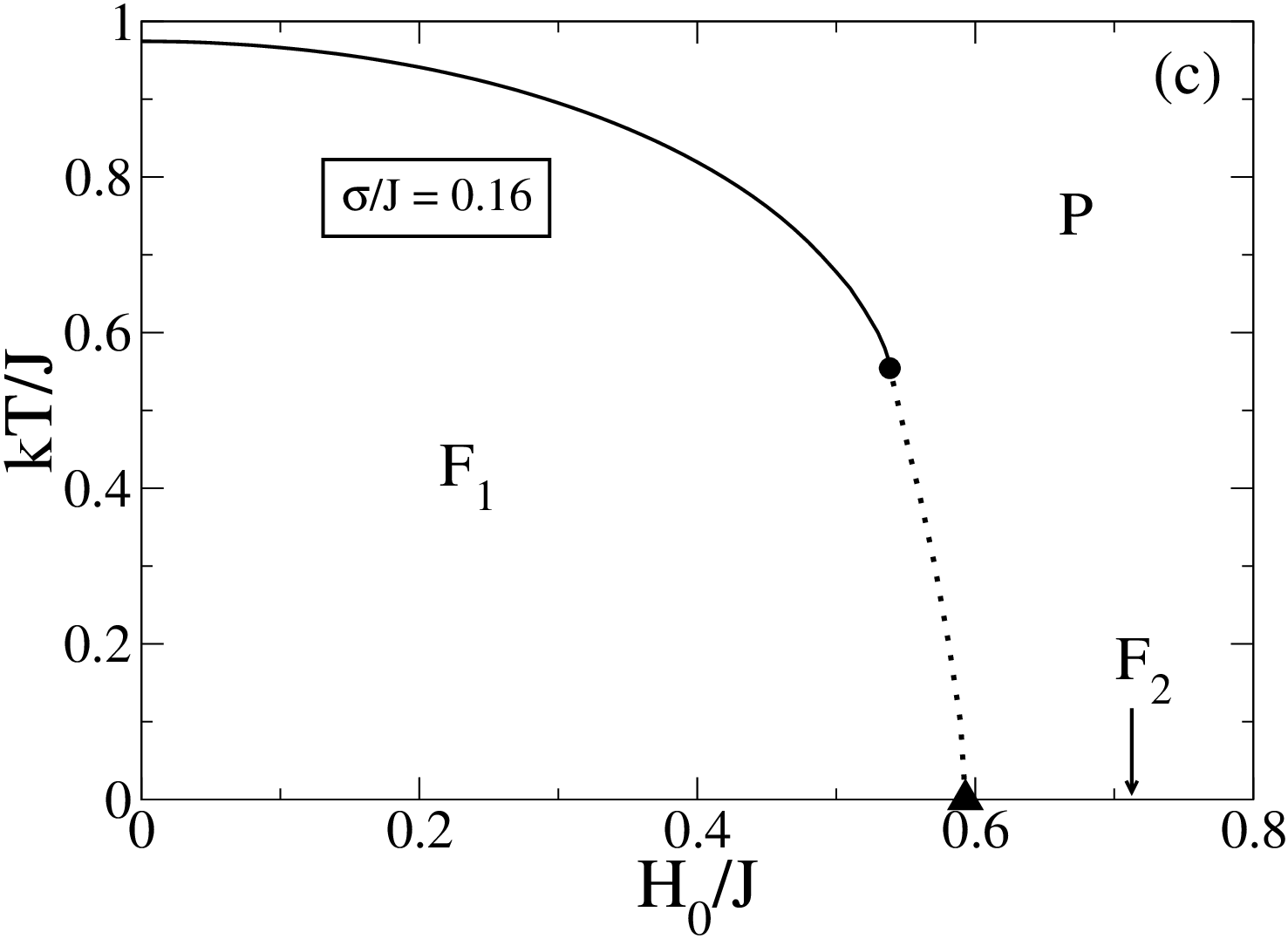}
\hspace{1.5cm}
\includegraphics[width=0.35\textwidth,angle=0]{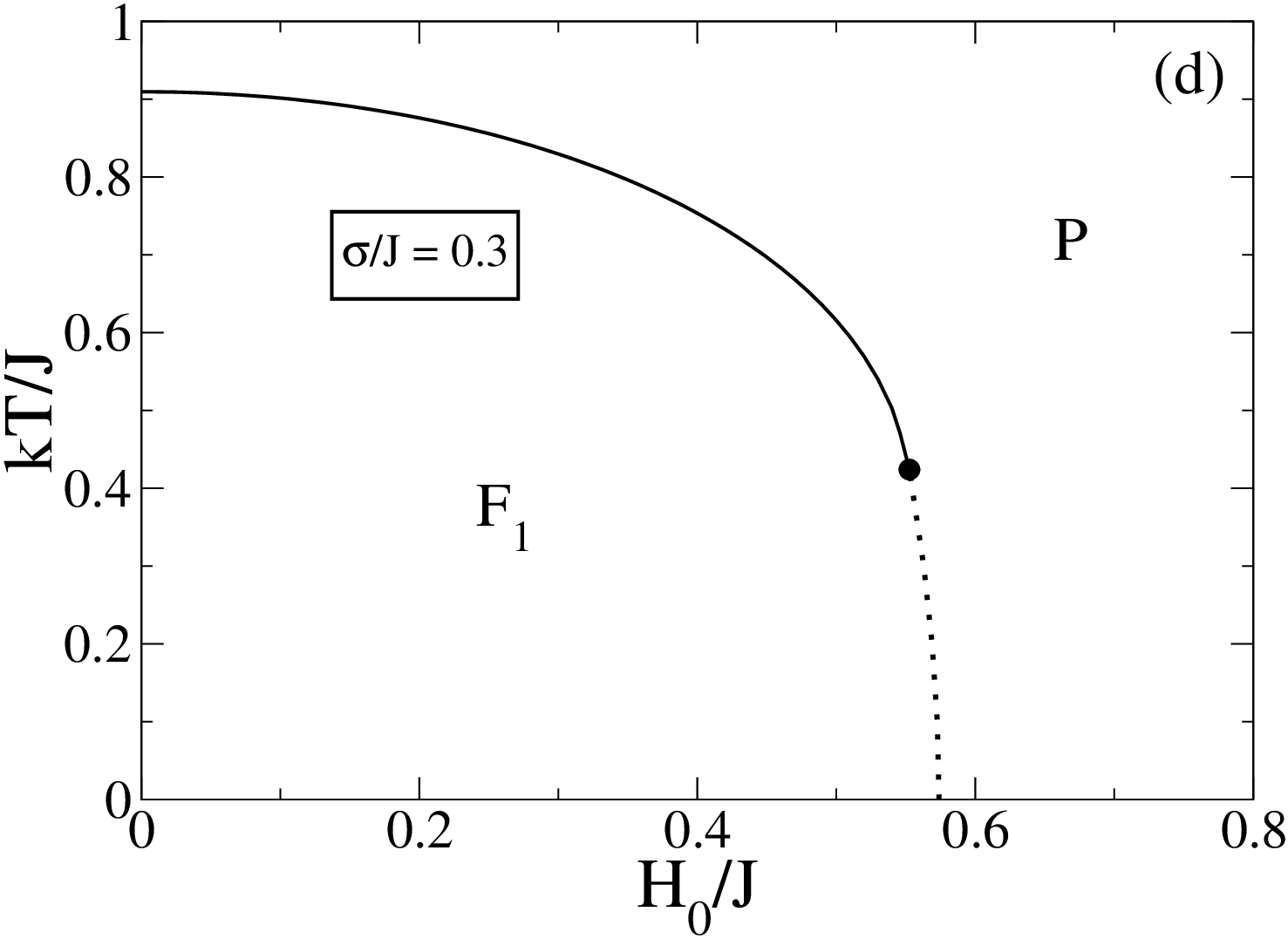}
\\
\vspace{1.0cm}
\includegraphics[width=0.35\textwidth,angle=0]{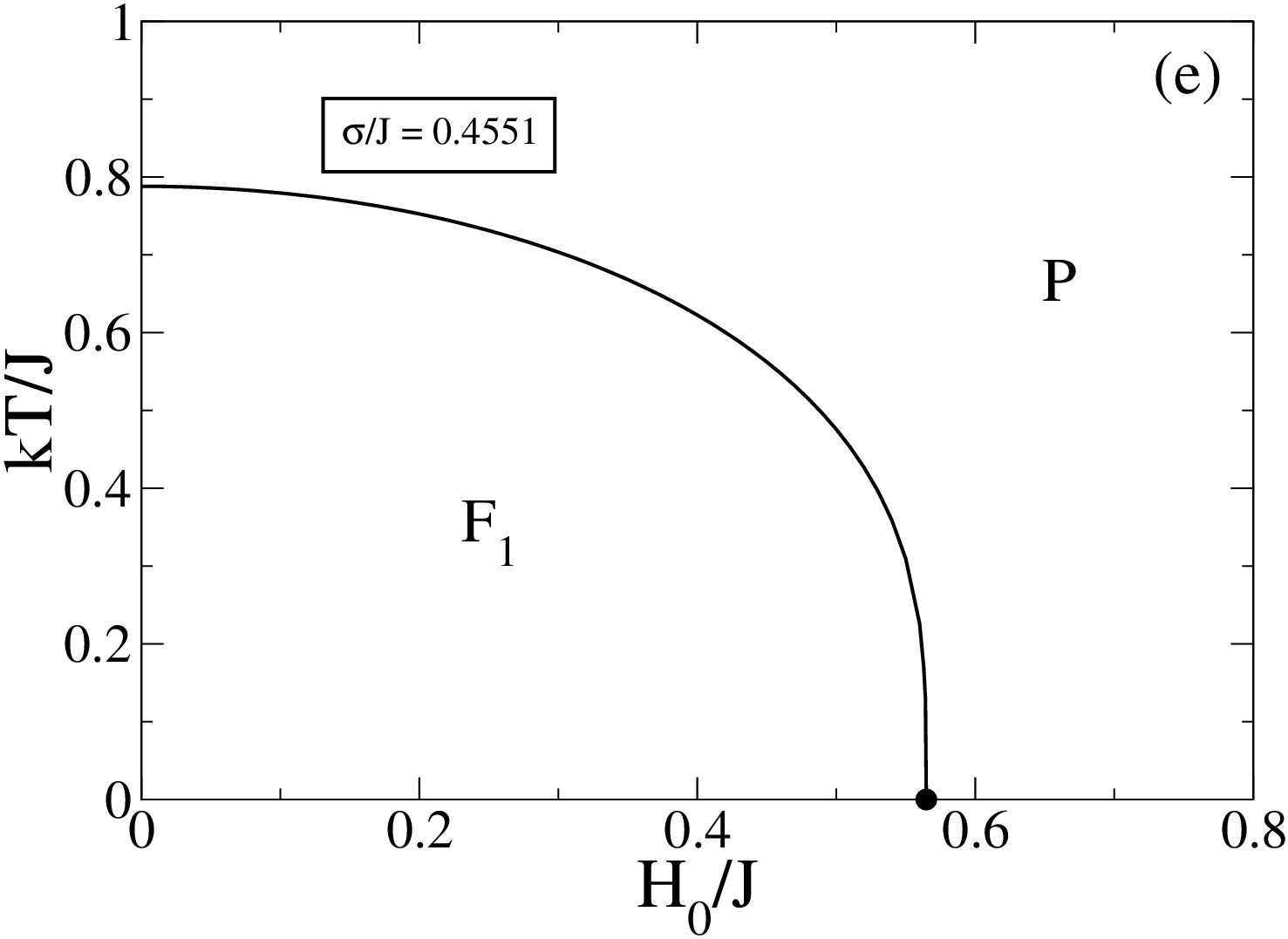}
\hspace{1.5cm}
\includegraphics[width=0.35\textwidth,angle=0]{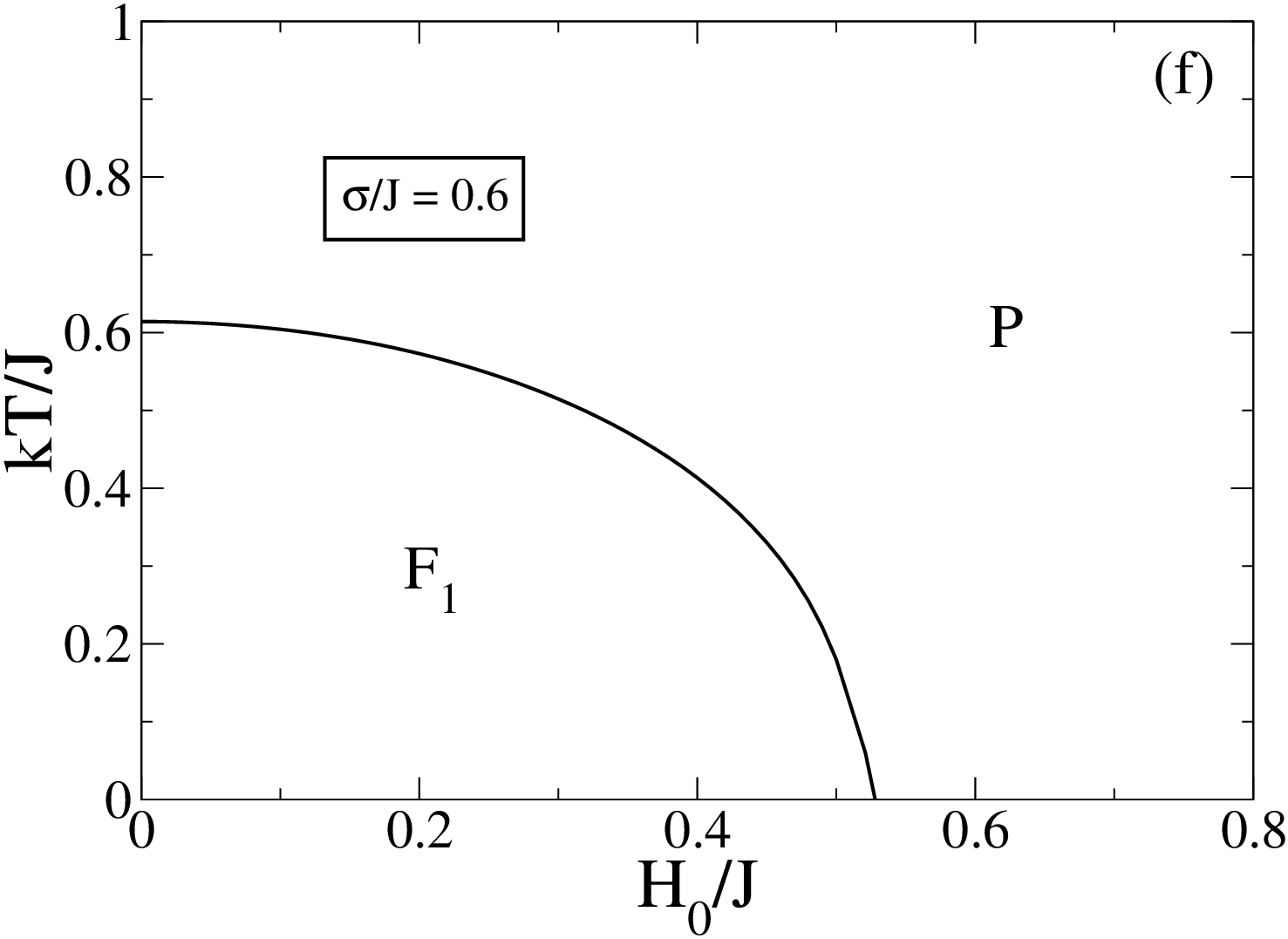}
\end{center}
\protect\caption{(a) The probability distribution of Eq.~(\ref{2}), for
$p=0.2$ and $(H_{0}/J)=0.56$ is represented for several values of 
its width. (b)--(f) Phase
diagrams showing the critical frontiers separating the paramagnetic ({\bf P})
and ferromagnetic ({\bf F}$\!_1$ and {\bf F}$\!_2$) phases of the
infinite-range-interaction ferromagnet in the presence of a triple Gaussian
random field, for typical values of $\sigma /J$ and $p=0.2$. Critical
frontiers and critical points are as described in the text. All quantities
are scaled in units of $J$.}
\label{fig1}
\end{figure}

In Fig.~\ref{fig1} we present some qualitatively distintic phase diagrams 
of the model, for the fixed value $p=0.2$ and typical values of $\sigma/J$. 
In Fig.~\ref{fig1}(a) we represent the PDF of  
Eq.~(\ref{2}) for $p=0.2$, $(H_{0}/J)=0.56$, and the widths 
$\sigma/J$ used to obtain the
phase diagrams of Figs.~\ref{fig1}(c)--(f). The particular case
$(\sigma/J)=0.1$, used in Fig.~\ref{fig1}(b), yields a 
three-narrow-peaked distribution and is not represented in 
Fig.~\ref{fig1}(a) for a better visualization of the remaining cases;
in addition to that, the choice $(H_{0}/J)=0.56$ corresponds to a
region in the phase diagrams where essential changes 
occur in the criticality of the system at low temperatures. 
One notices that the critical frontier separating the paramagnetic ({\bf P})
and ferromagnetic ({\bf F}$\!_1$ and {\bf F}$\!_2$) phases  
exhibits significant changes with increasing values of $\sigma/J$. 
For $(\sigma/J)=0.1$ [cf. Fig.~\ref{fig1}(b)], one has a phase diagram that is 
qualitatively similar to the one obtained in the case
of the trimodal distribution \cite{kaufman}, with a tricritical point
(black circle) and a
critical end point (black triangle) at a lower, but finite temperature,
where the critical 
frontier separating phases {\bf P} and {\bf F}$\!_2$ meets the first-order
line. 
However, for increasing values of
$\sigma/J$ these critical points move towards low temperatures, in such a way
that for
$(\sigma/J)=0.16$ one observes the collapse of the critical end point with
the zero-temperature axis, i.e., the ferromagnetic phase {\bf F}$\!_2$ 
occurs only for $T=0$, as shown in Fig.~\ref{fig1}(c). Therefore, 
for $p=0.2$ the value $(\sigma/J)=0.16$ represents a threshold for the
existence of phase {\bf F}$\!_2$. One should notice from  
Fig.~\ref{fig1}(a), that for $p=0.2$, 
the existence of the phase {\bf F}$\!_2$ is associated with a 
PDF for the fields characterized by a three-peaked shape. 
For $(\sigma/J)=0.3$ [cf. Fig.~\ref{fig1}(d)], the phase diagram is 
qualitatively similar to the one obtained for the bimodal \cite{aharony78}
or continuous two-peaked distributions \cite{nuno08a}, where one finds a
tricritical point at finite temperatures and a first-order transition at
lower temperatures. 
We have found analitically through a zero-temperature analysis that will be
discussed later,
the value $(\sigma/J) \cong 0.4551$ for which the 
tricritical point reaches the zero-temperature axis, 
signaling a complete
destruction of the first-order phase transition in the case $p=0.2$, 
as exhibited in 
Fig.~\ref{fig1}(e). In this case, the associated PDF presents a single flat
maximum in agreement with Refs.~\cite{aharony78,andelman,galambirman}. 
For even higher values of $\sigma/J$, as shown in 
Fig.~\ref{fig1}(f) for the case $(\sigma/J)=0.6$, the PDF presents a single
peak, and the frontier
ferromagnetic-paramagnetic is completely continuous, qualitatively similar
to the case of a
single Gaussian PDF \cite{schneiderpytte}. 

\begin{figure}[t]
\begin{center}
\includegraphics[width=0.35\textwidth,angle=0]{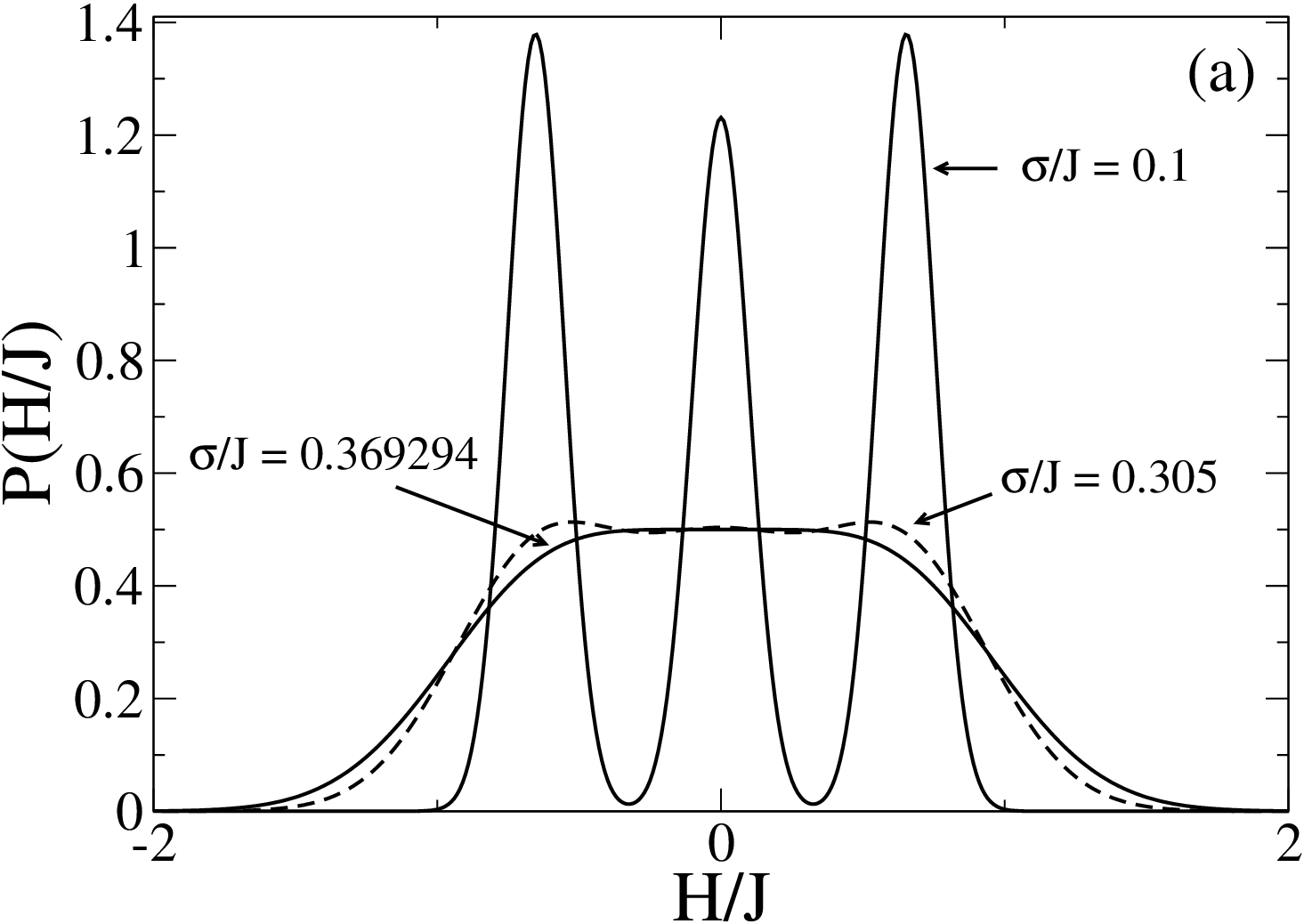}
\hspace{1.5cm}
\includegraphics[width=0.35\textwidth,angle=0]{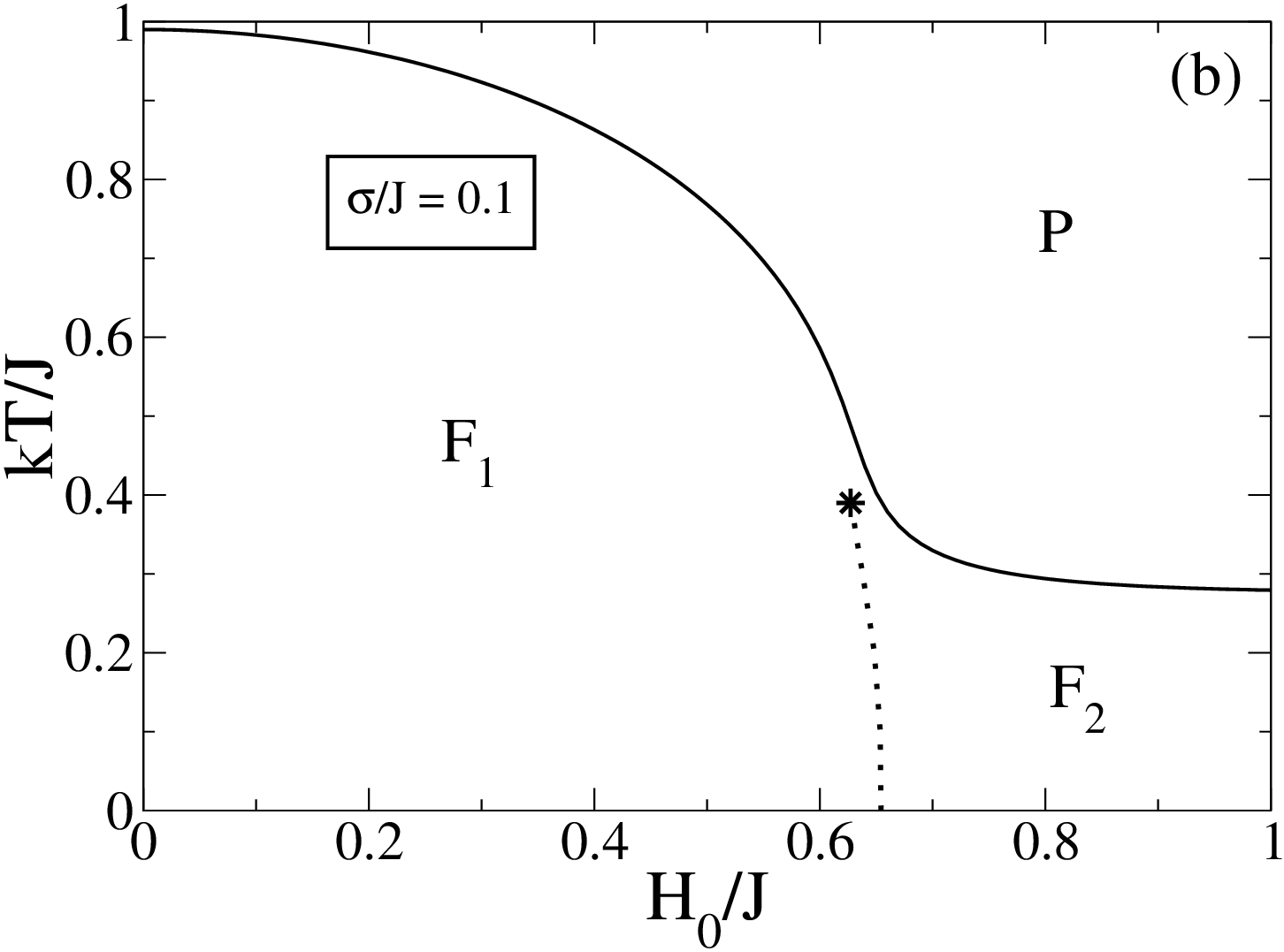}
\\
\vspace{1.0cm}
\includegraphics[width=0.35\textwidth,angle=0]{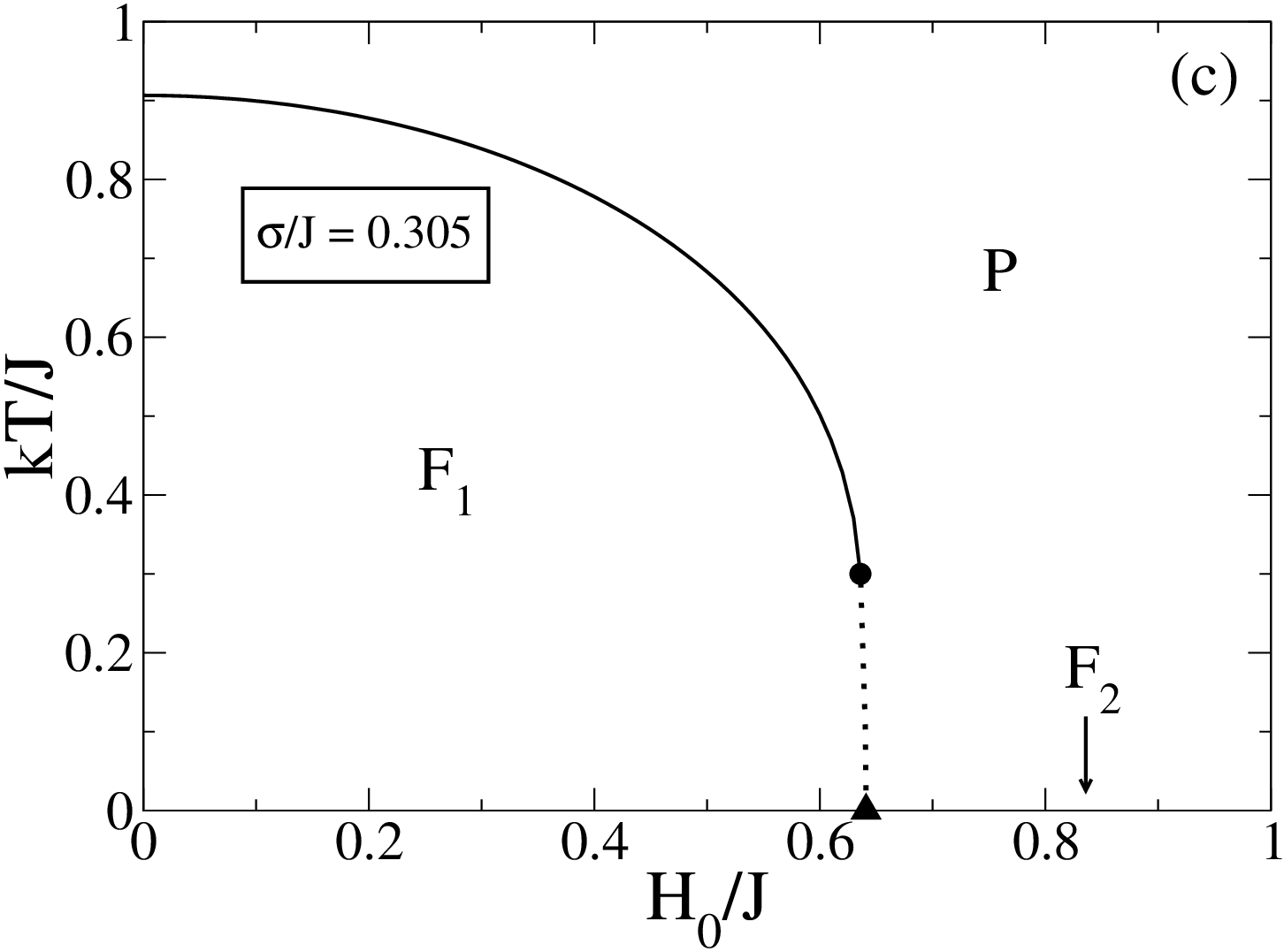}
\hspace{1.5cm}
\includegraphics[width=0.35\textwidth,angle=0]{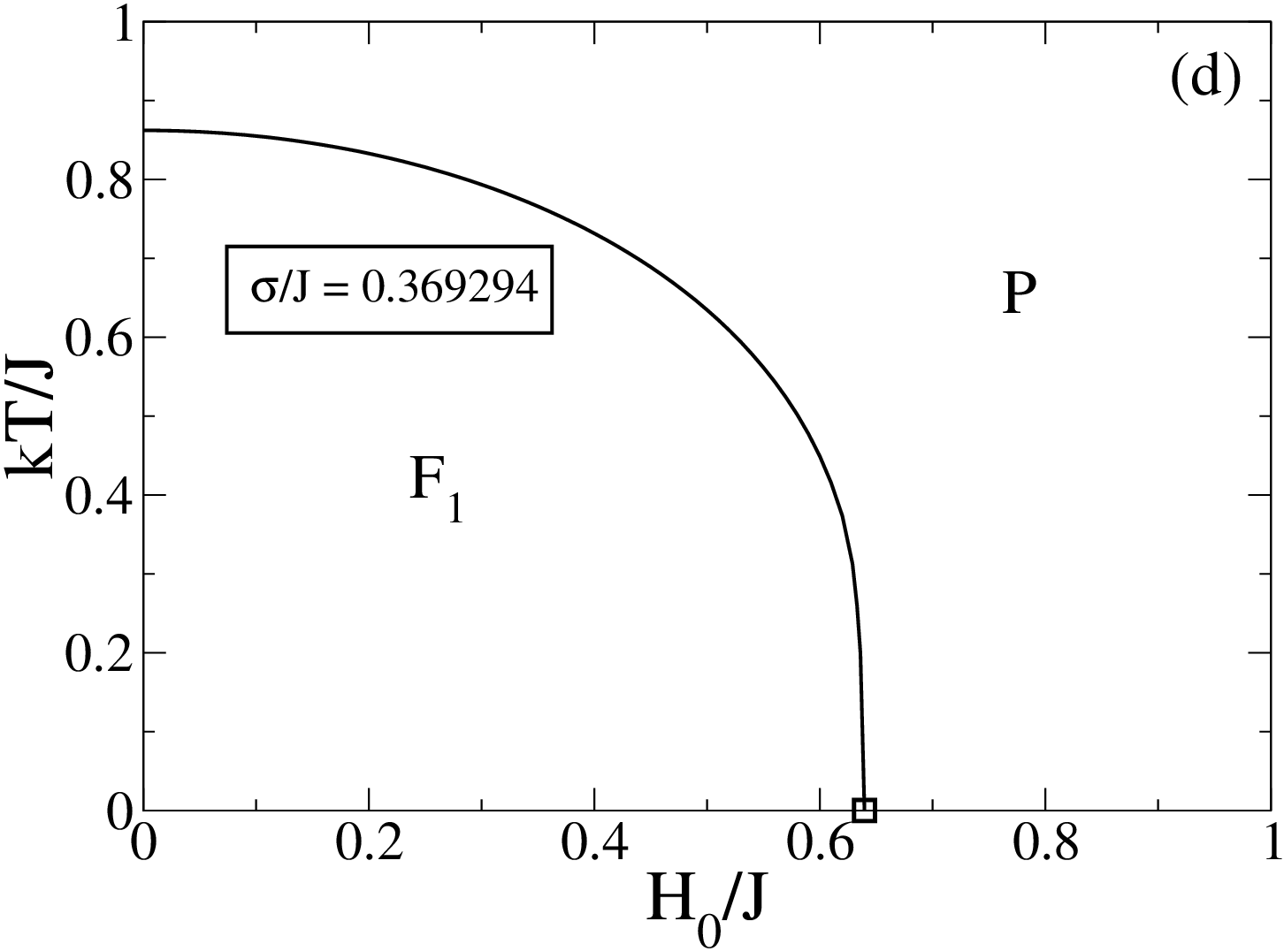}
\end{center}
\protect\caption{(a) The probability distribution of Eq.~(\ref{2}), for
$p=0.308561$ and $(H_{0}/J)=0.65$, is represented for 
several values of its width. 
(b)--(d) Phase
diagrams showing the critical frontiers separating the paramagnetic ({\bf P})
and ferromagnetic ({\bf F}$\!_1$ and {\bf F}$\!_2$) phases of the
infinite-range-interaction ferromagnet in the presence of a triple Gaussian
random field, for typical values of $\sigma /J$ and $p=0.308561$. Critical
frontiers and critical points are as described in the text. All quantities
are scaled in units of $J$.}
\label{fig2}
\end{figure}

There is an special value of $p$, to be denoted herein $p^{*}$, 
which represents an upper limit for the existence of tricritical points
along the paramagnetic border, characterized by a 
fourth-order point at zero temperature. The value 
$p^{*} \cong 0.308561$, as well as its associated width, 
$(\sigma^{*}/J) \cong 0.369294$,  
will be determined later, within a zero-temperature
analysis. 
In Fig.~\ref{fig2} we show phase diagrams for $p=0.308561$  
and three different values of $\sigma/J$. 
In Fig.~\ref{fig2}(a) we represent the PDF of  
Eq.~(\ref{2}) for $p=0.308561$, $(H_{0}/J)=0.65$, and the 
widths $\sigma/J$ used to obtain the
phase diagrams of Figs.~\ref{fig2}(b)--(d); 
the choice $(H_{0}/J)=0.65$ has to do with a
region in the phase diagrams where interesting 
critical phenomena occur at low temperatures. 
In Fig.~\ref{fig2}(b) we present the phase diagram for $(\sigma/J)=0.1$,
where one sees that the border of
the paramagnetic phase is completely continuous; for lower (higher) fields 
this critical frontier separates phases {\bf P} and 
{\bf F}$\!_1$ ({\bf F}$\!_2$).
However, there is a curious 
first-order line separating phases {\bf F}$\!_1$ and {\bf F}$\!_2$ that
terminates in an ordered critical point (black asterisk), 
above which one can pass 
smoothly from one of these phases to the other. 
By slightly increasing the values of $\sigma/J$ one observes that the
ordered critical point disappears, leading to the emergence of both 
tricritical and critical end points at finite temperatures, 
yielding a phase diagram that is 
qualitatively similar to the one shown in Fig.~\ref{fig1}(b); this type of
phase diagram occurs
typically in the range $0.1 < (\sigma/J) < 0.3$.
However, as shown in Fig.~\ref{fig2}(c) for $(\sigma/J)=0.305$, one has 
the collapse of the critical end point with the zero-temperature axis, 
with the phase {\bf F}$\!_2$ occurring at zero temperature only, 
analogous to the one that appears in Fig.~\ref{fig1}(c), but now for a
higher value of $H_{0}/J$. Comparing Figs.~\ref{fig1}(c) and \ref{fig2}(c),
one sees that in some cases, essentially similar phase diagrams may be
obtained by increasing both 
$p$ and $\sigma/J$; however, in the case of higher values for these 
parameters, the extension of the
first-order transition line gets reduced. It should be mentioned that, in
both figures,  the existence of the phase {\bf F}$\!_2$ is associated with
a PDF for the fields characterized by a three-peaked shape.
In Fig.~\ref{fig2}(d) we display the phase diagram for  
$(\sigma/J) = 0.369294$, for which the first-order line is totally
destroyed. It should be emphasized that this effect appears also  
for other values of $p$ through a tricritical point at $T=0$ 
[cf. Fig.~\ref{fig1}(e)]; however, the phase diagram of Fig.~\ref{fig2}(d)
is very special, in the sense that    
the collapse of the first-order frontier with the zero-temperature axis
occurs by means of a fourth-order point.   
For higher values of $\sigma/J$, the critical frontier separating phases
{\bf P} and {\bf F}$\!_1$ is completely continuous, and one has essentially
the same phase diagram shown in Fig.~\ref{fig1}(f). 

\begin{figure}[t]
\begin{center}
\includegraphics[width=0.35\textwidth,angle=0]{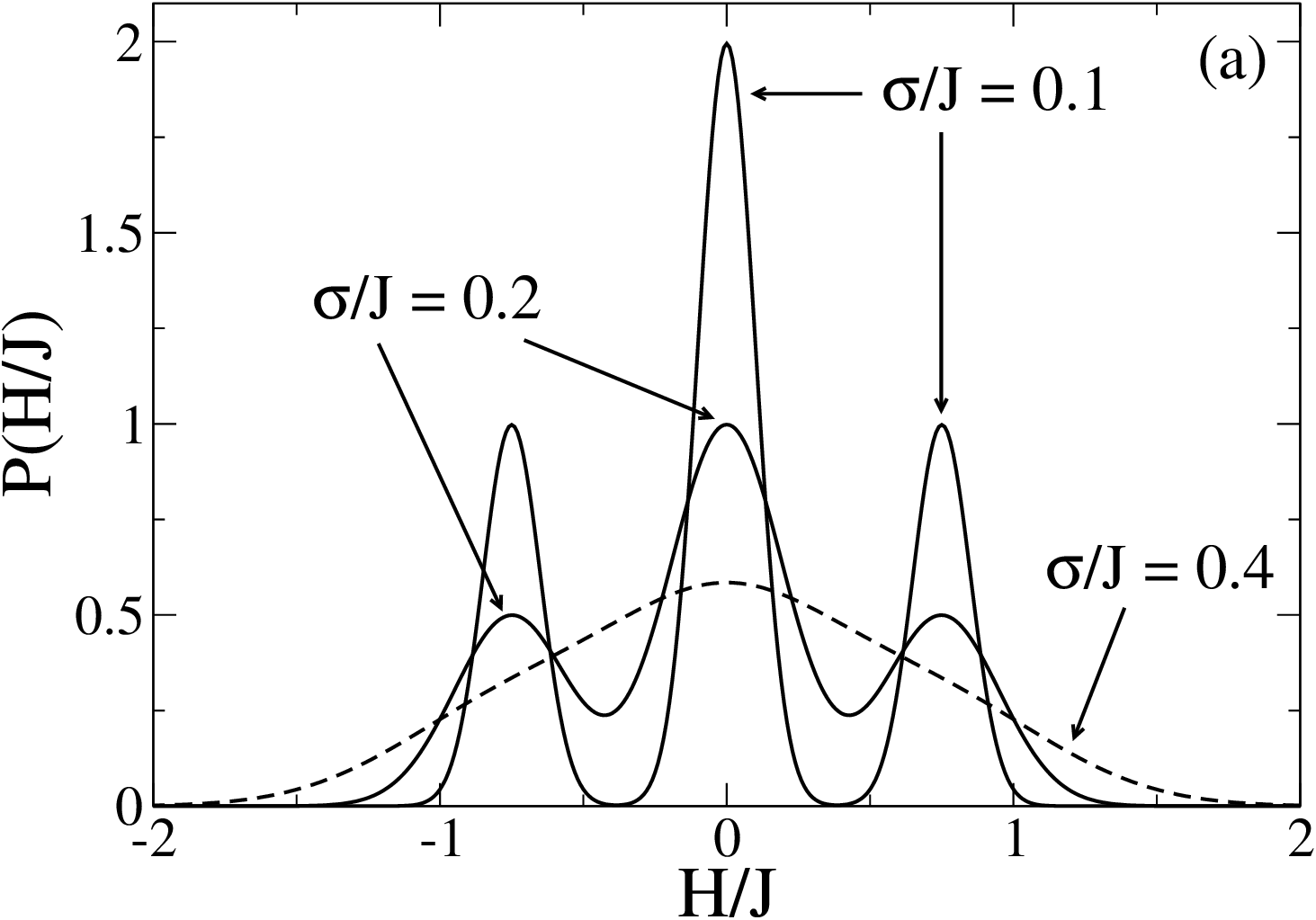}
\hspace{1.5cm}
\includegraphics[width=0.35\textwidth,angle=0]{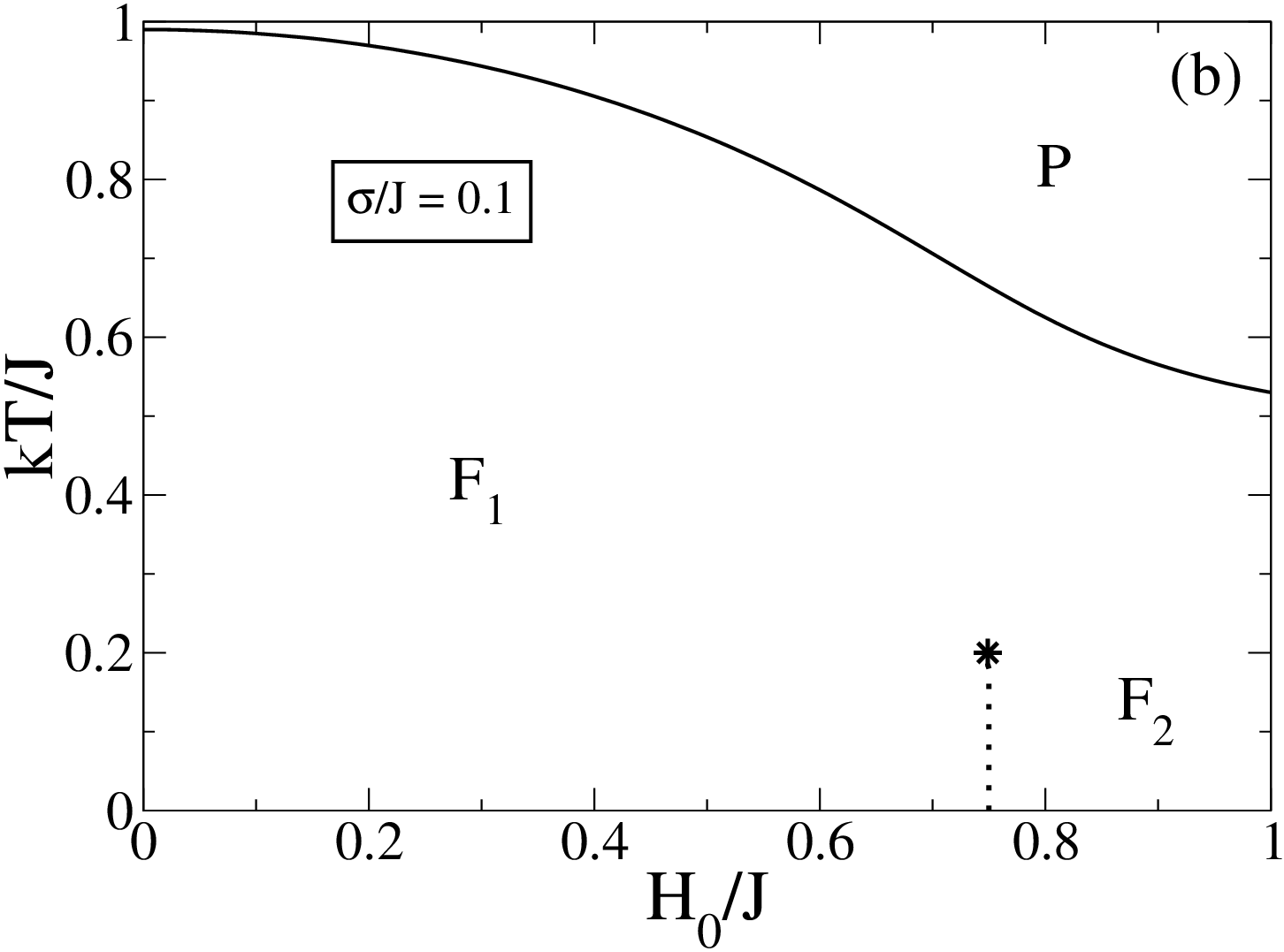}
\\
\vspace{1.0cm}
\includegraphics[width=0.35\textwidth,angle=0]{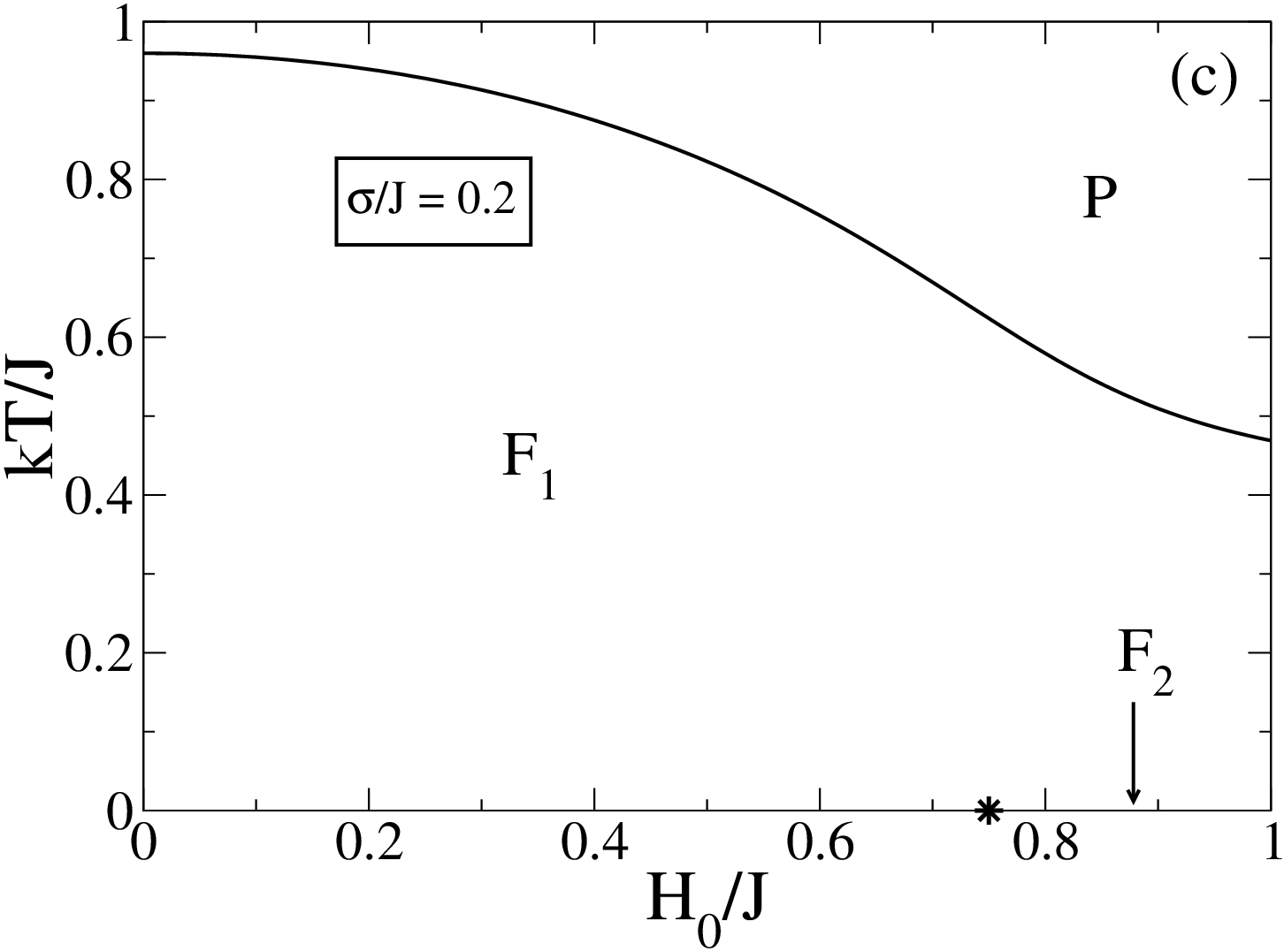}
\hspace{1.5cm}
\includegraphics[width=0.35\textwidth,angle=0]{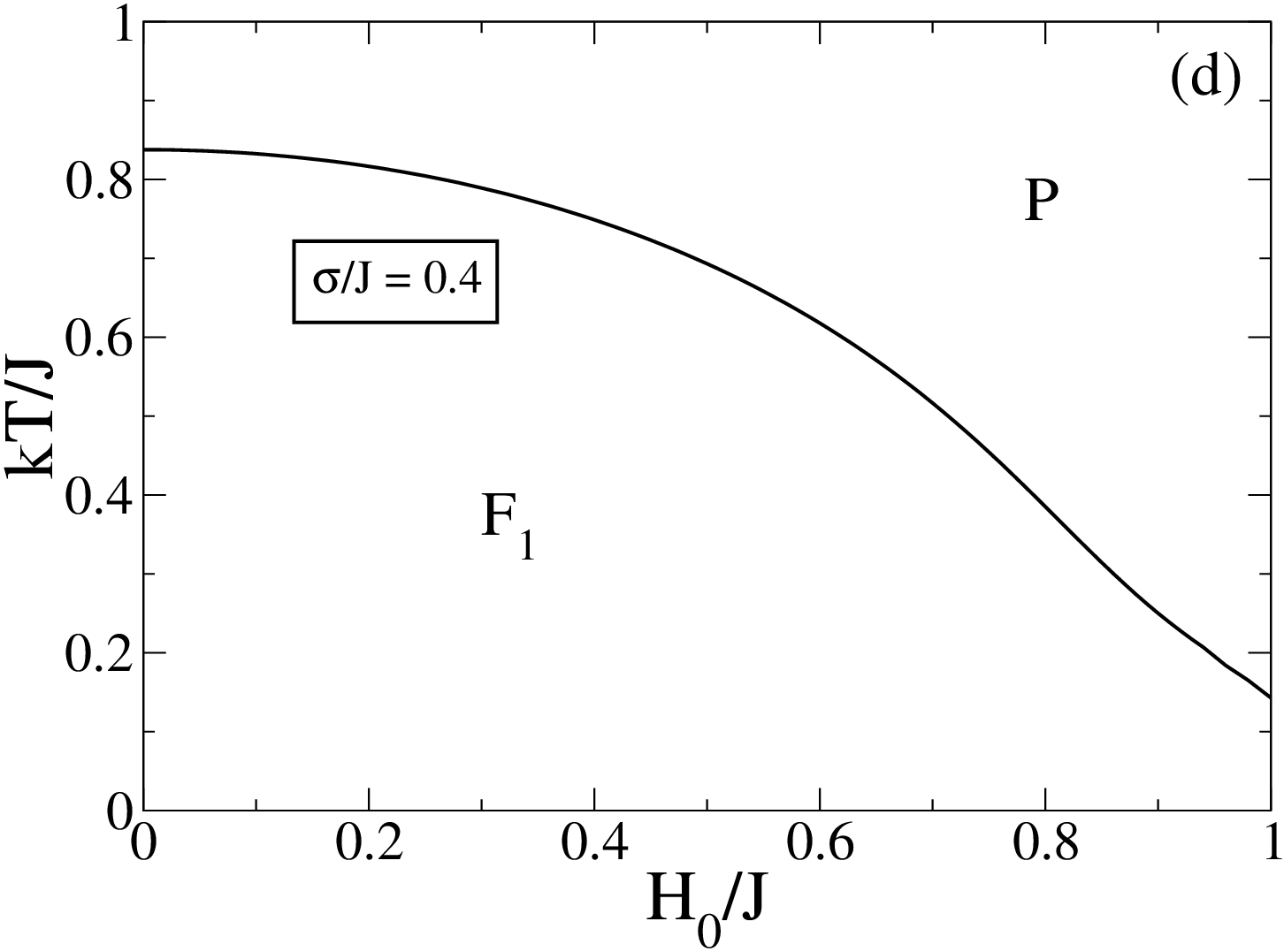}
\end{center}
\protect\caption{(a) The probability distribution of Eq.~(\ref{2}), for
$p=0.5$ and $(H_{0}/J)=0.74$, is represented for 
several values of its width. 
(b)--(d) Phase
diagrams showing the critical frontiers separating the paramagnetic ({\bf P})
and ferromagnetic ({\bf F}$\!_1$ and {\bf F}$\!_2$) phases of the
infinite-range-interaction ferromagnet in the presence of a triple Gaussian
random field, for typical values of $\sigma /J$ and $p=0.5$. Critical
frontiers and critical points are as described in the text. All quantities
are scaled in units of $J$.}
\label{fig3}
\end{figure}

Additional phase diagrams are shown in Fig.~\ref{fig3}, for the case
$p=0.5$. 
In Fig.~\ref{fig3}(a) we represent the PDF of  
Eq.~(\ref{2}) for such a value of $p$, $(H_{0}/J)=0.74$, and the 
widths $\sigma/J$ used to obtain the
phase diagrams of Figs.~\ref{fig3}(b)--(d); 
the choice $(H_{0}/J)=0.74$ corresponds to a region of the phase diagram
where an ordered critical point appears at low temperatures.  
In Fig.~\ref{fig3}(b) we present the phase diagram for the same 
$\sigma/J$ value of Figs.~\ref{fig1}(b) and \ref{fig2}(b), i.e., 
$(\sigma/J)=0.1$; these figures suggest that the existence of 
phase {\bf F}$\!_2$ is associated 
with three-peaked distributions. However, for sufficiently small $p$, this
phase appears together with a critical end point, as shown in
Fig.~\ref{fig1}(b), whereas for larger $p$, the border of the paramagnetic
phase is completely continuous, and phases {\bf F}$\!_1$  and {\bf F}$\!_2$
are separated by a first-order critical frontier that terminates in an
ordered critical point.   
In Fig.~\ref{fig3}(b) one can go smoothly from one of these two
ferromagnetic phases to the other, through a thermodynamic path connecting
these phases above the ordered critical point.   
In Fig.~\ref{fig3}(c) one has the collapse of the ordered critical point
with the zero-temperature axis, which was estimated numerically to occur
for  
$(\sigma/J)=0.20$, with the phase {\bf F}$\!_2$ appearing at zero temperature. 
It is important to mention that the phase diagram exhibited in  
Fig.~\ref{fig3}(c) is qualitatively distinct from those of 
Figs.~\ref{fig1}(c) and \ref{fig2}(c), in the sense that the first one is
characterized by a border of the paramagnetic phase that is completely
continuous and at zero temperature, phases {\bf F}$\!_1$  and {\bf F}$\!_2$
are separated by  
an ordered critical point.
For the value $(\sigma/J)=0.4$ shown in 
Fig.~\ref{fig3}(d), one gets a continuous critical frontier separating 
phases {\bf P} and {\bf F}$\!_1$. However, there is a basic difference
between the phase diagrams displayed in   
Figs.~\ref{fig1}(f) and \ref{fig3}(d): as it will be shown in the following
zero-temperature analysis, in the later case there is no zero-temperature
point separating these two phases, i.e., the phase {\bf F}$\!_1$ exists for
all values  
of $H_{0}/J$. In fact, for any 
$(\sigma/J) \leq 0.4$ the border of the paramagnetic phase never touches the
zero-temperature axis, as suggested in Figs.~\ref{fig3}(b)--(d);  
for higher values of $\sigma/J$, this 
critical line meets the $H_{0}/J$ axis, and one has a phase
diagram qualitatively similar to the one of Fig.~\ref{fig1}(f).

\begin{figure}[t]
\begin{center}
\includegraphics[width=0.35\textwidth,angle=0]{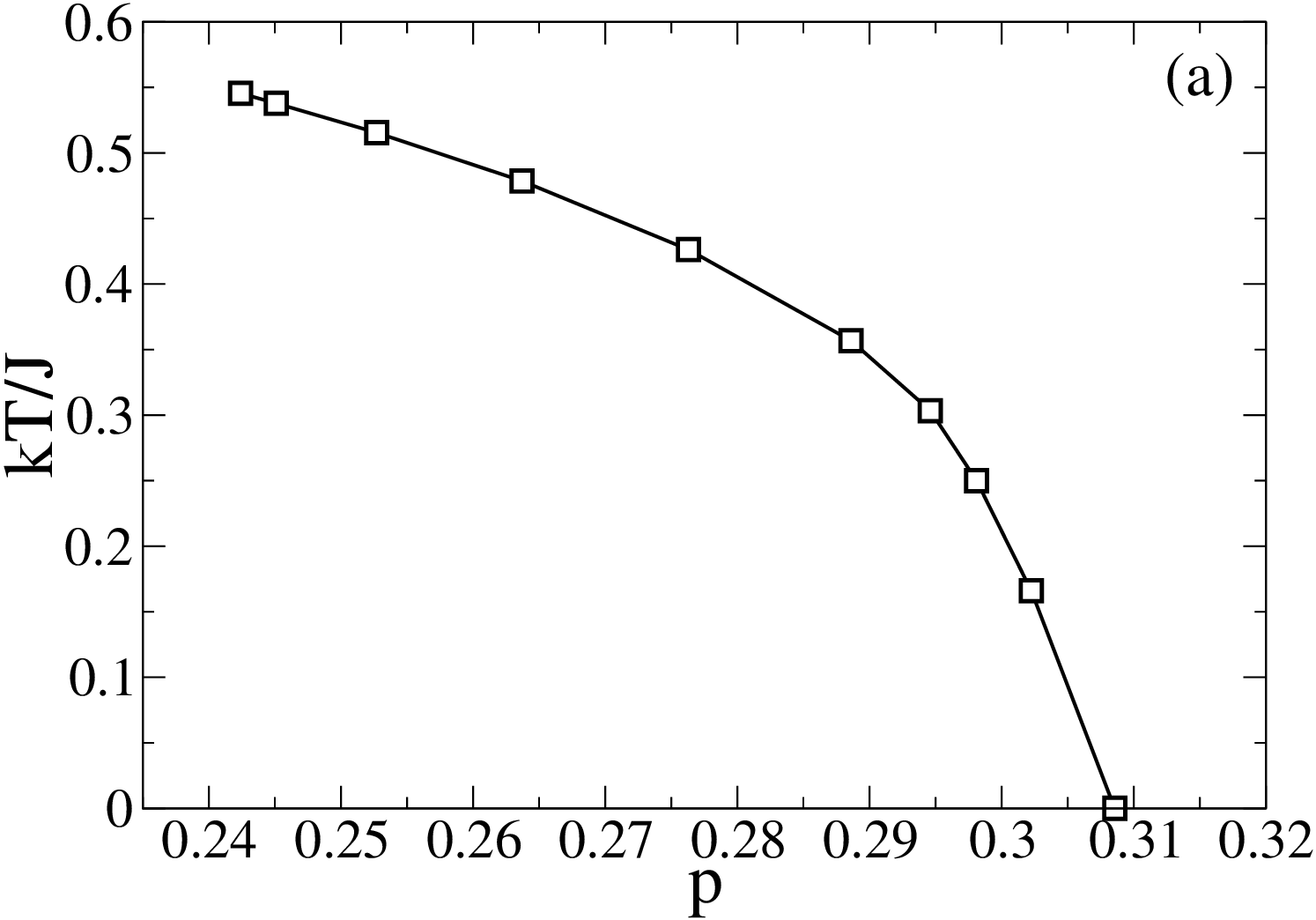}
\hspace{1.5cm}
\includegraphics[width=0.35\textwidth,angle=0]{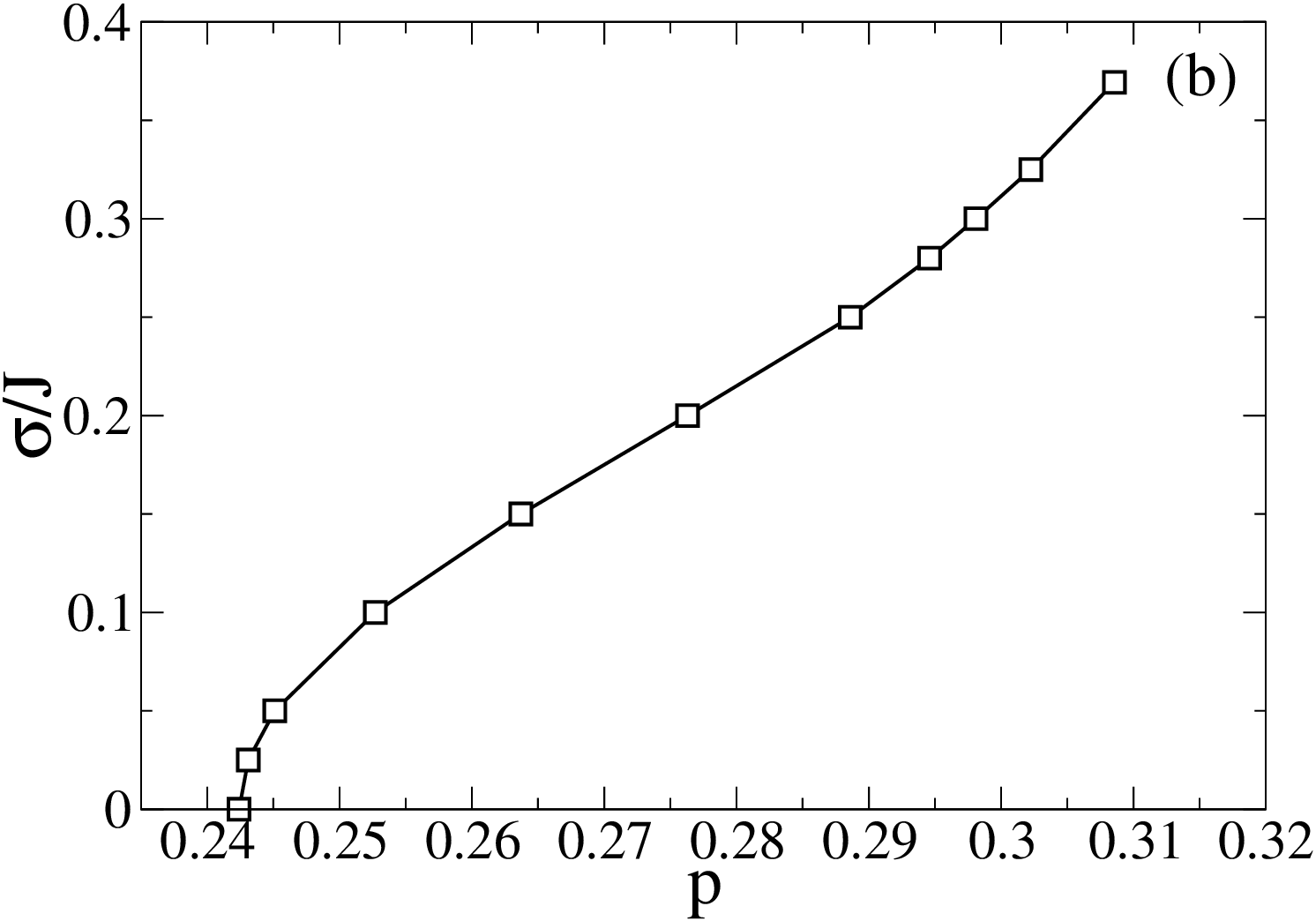}
\end{center}
\protect\caption{Projections of the fourth-order-point line on two
different planes:  
(a)  The plane $(kT)/J$ versus $p$, showing the critical temperatures
associated with fourth-order points that exist in the interval   
$8/33 \leq p \leq p^{*}$; 
(b) The plane $\sigma/J$ versus $p$. The empty squares are computed
fourth-order points, whereas the lines (guides to the eye) represent lines
of fourth-order points.} 
\label{fig4}
\end{figure}

There are some important threshold values associated with the existence 
of the above-mentioned critical points, as described next. 

(i) We found numerically that for $p \gtrsim p^{*}$, critical end points do
not occur, either in finite or zero temperatures. 

(ii) We verified, also numerically, that for $p \gtrsim 0.93$, ordered
critical points cease to exist; as a consequence, this represents a
threshold for the existence of phase {\bf F}$\!_2$, which does not appear
above this value (even at zero temperature).  

(iii) In the zero-temperature analysis (to be carried below), we find
analytically that for  $p^{*} \cong 0.308561$ one gets a fourth-order
point at zero temperature; this value depicts an upper bound for the
existence of tricritical points.  

(iv) Fourth-order points usually delimitate the existence of tricritical
points and are sometimes considered in the literature as ``vestigial''
tricritical points  
\cite{kaufman}. 
In the case of the power expansion in Eq.~(\ref{11}), they exist for finite
temperatures as well, and are determined by the conditions  
$A_{1}=1$, $A_{3}=A_{5}=0$ and $A_{7}<0$, which define  
a line in the four-dimensional space $[(kT)/J,p,H_{0}/J,\sigma/J]$. 
In Fig.~\ref{fig4} we present projections of the fourth-order-point
line with the planes $(kT)/J$ versus $p$ [Fig.~\ref{fig4}(a)] and
$\sigma/J$ versus $p$  
[Fig.~\ref{fig4}(b)]. These projections interpolate between the
fourth-order points occurring for $p=8/33$ ($\sigma=0$, i.e., trimodal
distribution \cite{kaufman}) and the zero-temperature threshold value for
the triple Gaussian PDF, $p^{*} \cong 0.308561$, to be determined below.

\begin{figure}[t]
\begin{center}
\includegraphics[width=0.35\textwidth,angle=0]{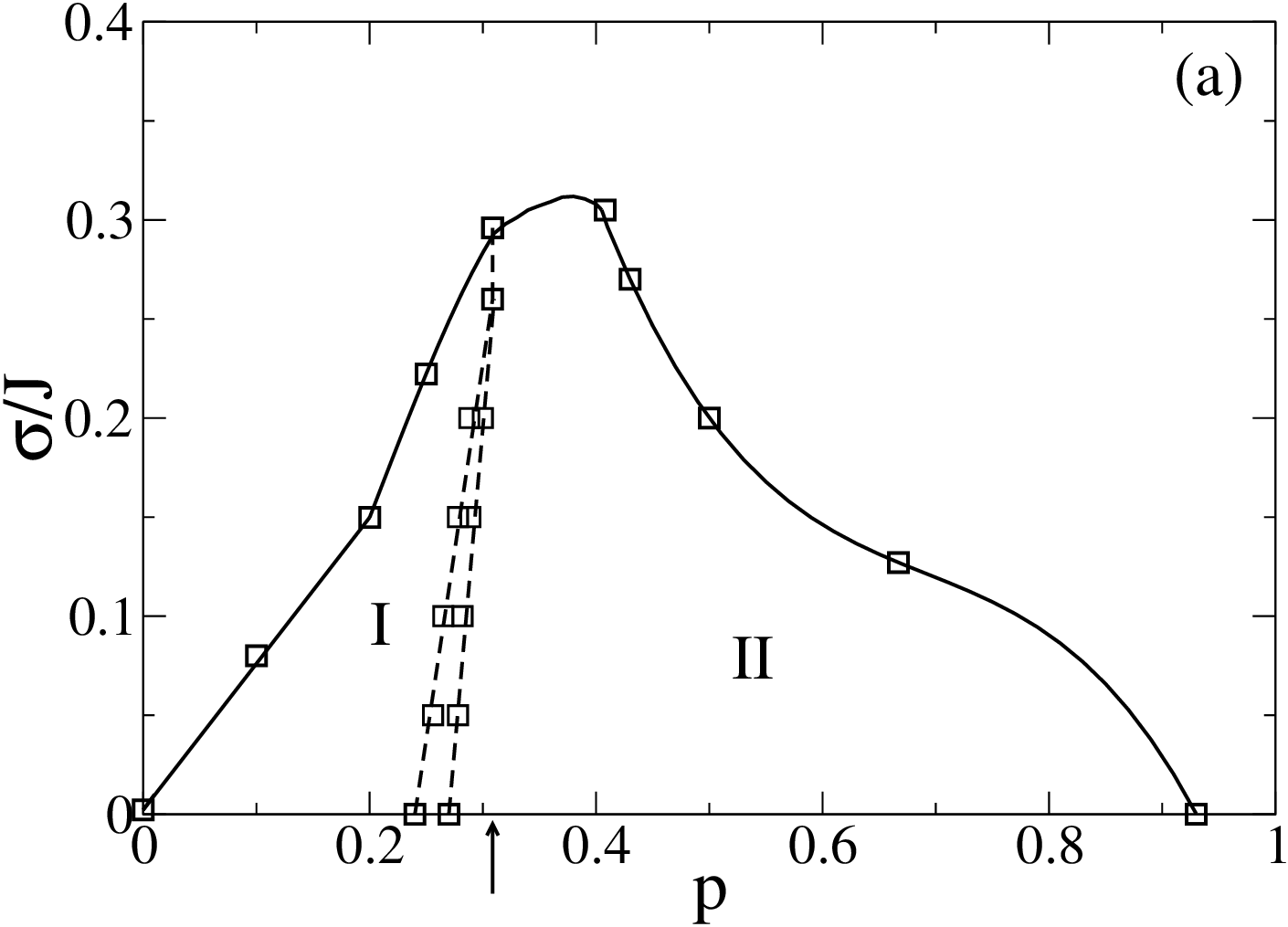}
\hspace{1.5cm}
\includegraphics[width=0.35\textwidth,angle=0]{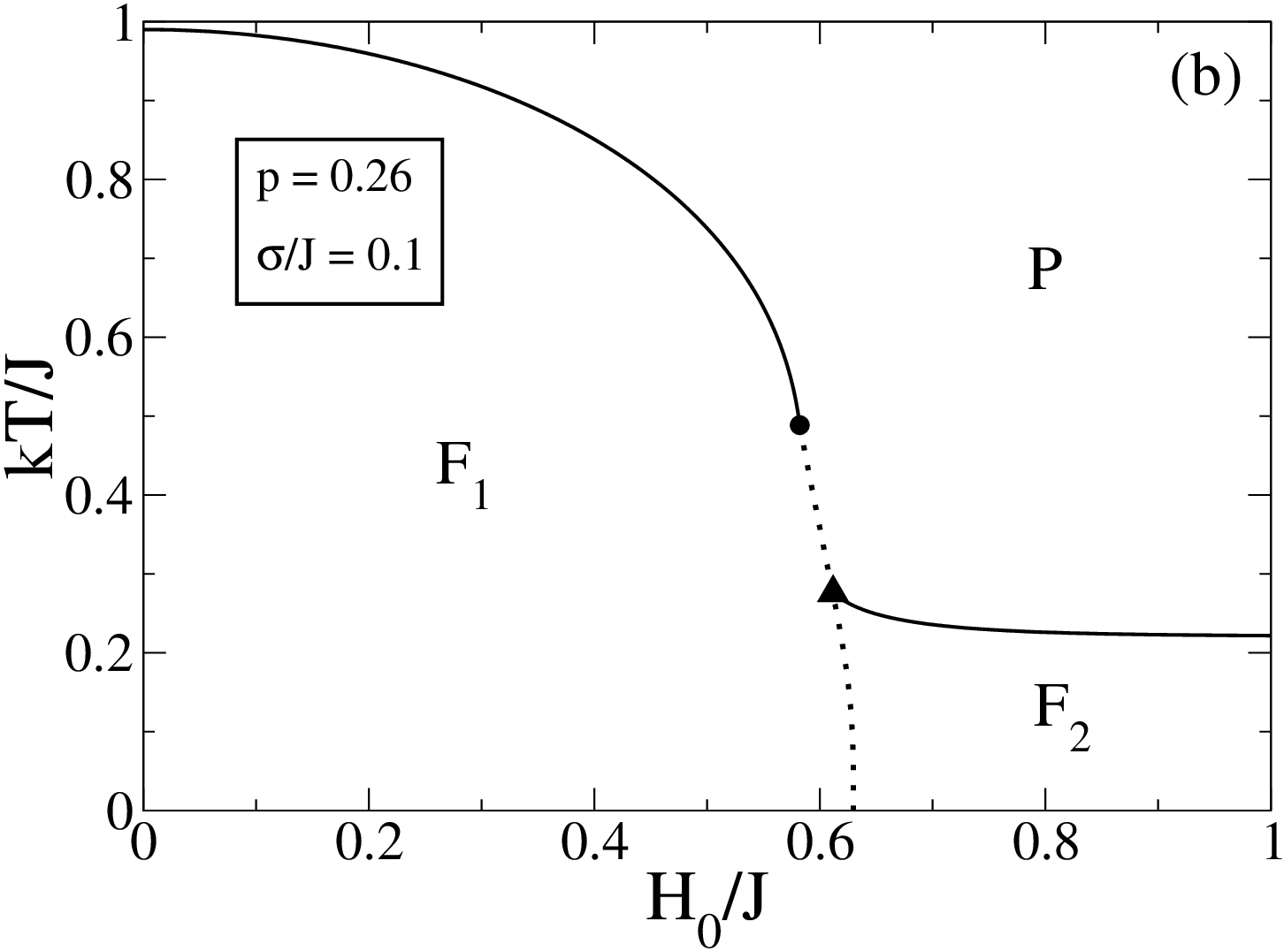}
\\
\vspace{1.0cm}
\includegraphics[width=0.35\textwidth,angle=0]{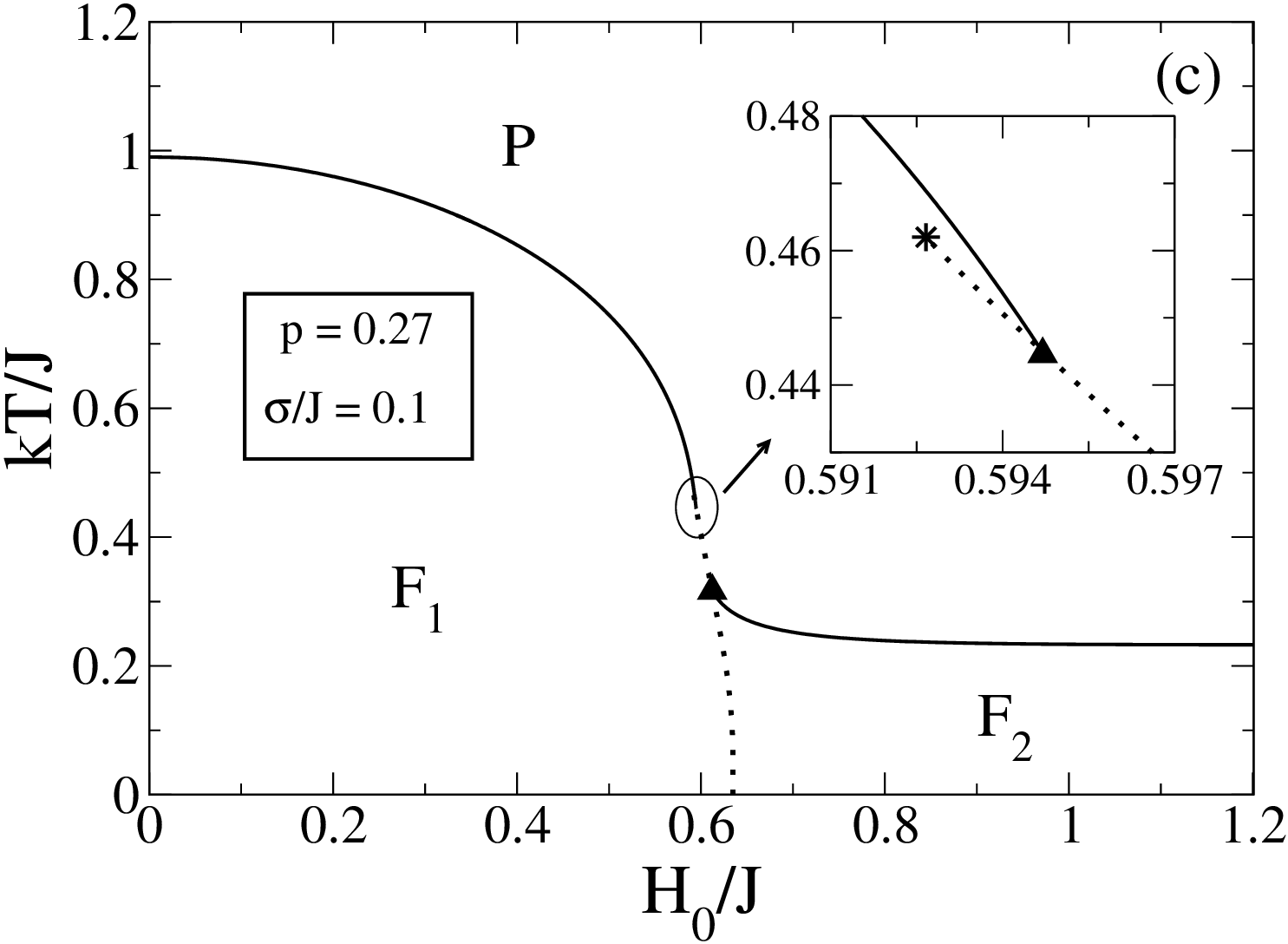}
\hspace{1.5cm}
\includegraphics[width=0.35\textwidth,angle=0]{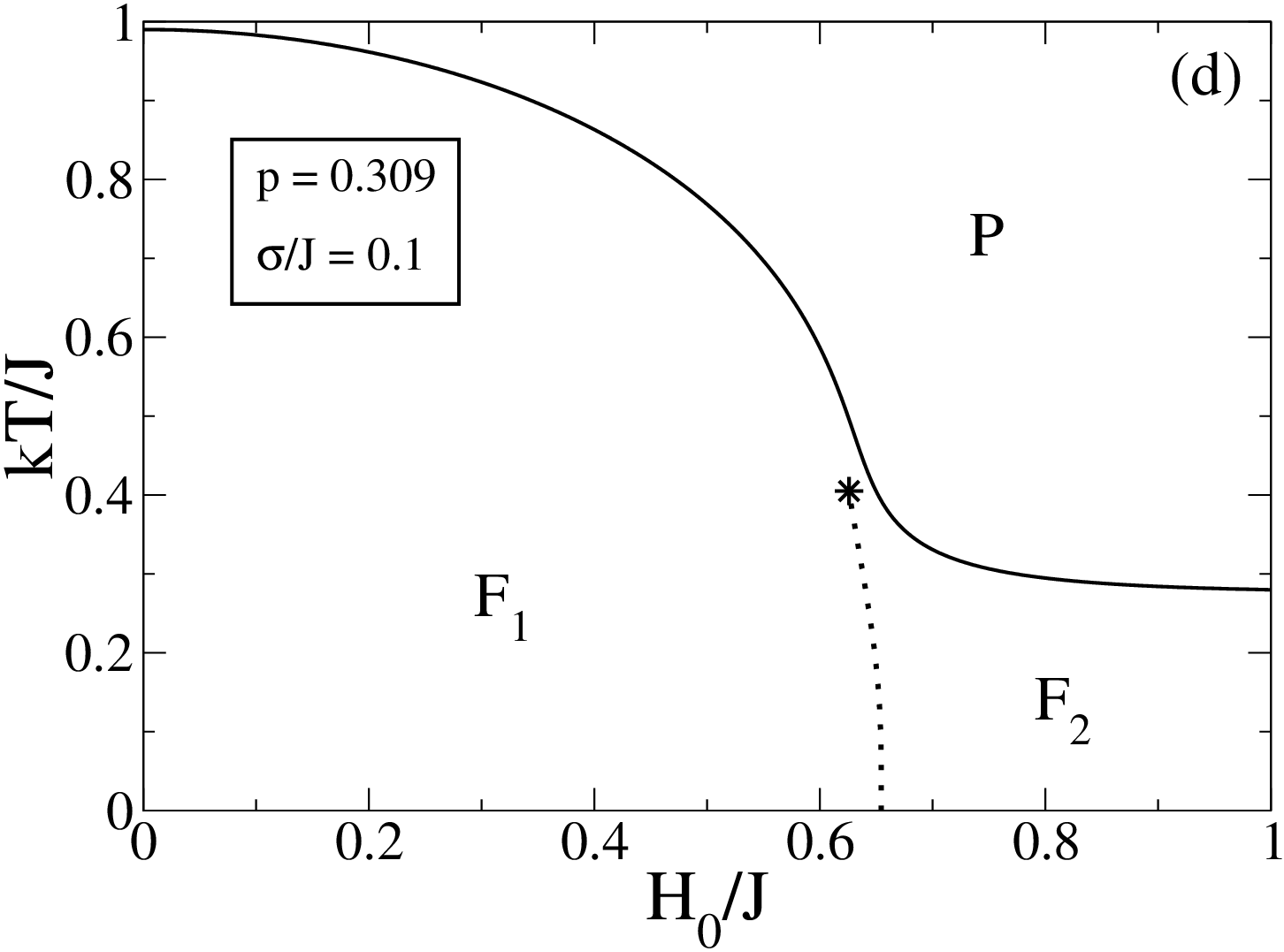}
\end{center}
\protect\caption{(a) Regions, in the plane $\sigma/J$ versus $p$, associated
with qualitatively distinct phase diagrams for the present model;  
the empty squares represent computed points, 
whereas the lines are just guides to the eye; the arrow indicates the
threshold value $p=p^{*}$.
In regions I and II one has phase diagrams like those shown
in (b) and (d), respectively. 
The two dashed lines that appear in (a) define a narrow intermediate 
region, exhibiting a phase diagram characterized by 
two critical end points and one ordered critical 
point along the paramagnetic border, like shown in (c). 
The type of
phase diagram in the intermediate region appears when one varies the
parameters of the distribution of Eq.~(\ref{2}) in such a way to go  
from region I to
region II [as shown typically in the sequence (b), (c) and (d), for
$(\sigma/J)=0.1$], and vice versa.}
\label{fig5}
\end{figure}

From the analysis above one concludes that it is possible to obtain
qualitatively similar phase diagrams for different pairs of parameters 
$(p,\sigma/J)$. Essentially, these phase diagrams are defined by the
presence of the different types of critical points that may appear in this 
model. The existence of tricritical points was already discussed in 
items (iii) and (iv) above, whereas the regions in the plane
$\sigma/J$ versus $p$ associated with ordered and critical end points are
exhibited in Fig.~\ref{fig5}(a).
Along the axis $(\sigma/J)=0$ our limits for the existence of critical end
points ($0 < p \lesssim 0.24$) and ordered critical points  
($0.27 \lesssim p \lesssim 0.93$) are in agreement with those estimated in
Ref.~\cite{kaufman}; in between these two regimes, a small region occurs,  
which is very 
subtle, from the numerical point of view, where   
two critical end points and one ordered critical point show up. 
This intermediate region gets reduced for increasing values of 
$\sigma/J$ and is delimited by the two dashed lines shown in 
Fig.~\ref{fig5}(a).
A typical phase diagram inside this region is shown in Fig.~\ref{fig5}(c),
which is characterized by a
critical frontier separating phases {\bf P} and {\bf F}$\!_1$ that presents a
continuous piece, followed by a first-order line at lower temperatures, with
no tricritical point. It is important to stress that the type of phase
diagram of Fig.~\ref{fig5}(c) occurs in the particular case 
$(\sigma/J)=0$
for a range of values of $p$ right above $p=8/33 \approx 0.24$, 
which represents 
the probability at which a fourth-order point occurs at finite
temperatures~\cite{kaufman}. We have verified that a similar effect occurs
for finite values of $\sigma/J$, in the sense that the narrow 
intermediate region of Fig.~\ref{fig5}(a) corresponds to a region of 
points to the right of the projection of the fourth-order line in the plane 
$\sigma/J$ versus $p$ of Fig.~\ref{fig4}(b). This intermediate region
occurs whenever the fourth order point appears for finite temperatures and
it becomes essentially undiscernible, from the computational point of view,
as one approaches the threshold $p=p^{*}$, corresponding to the fourth-order 
point at zero temmperature; 
we verified numerically that 
this region disappears completely at the threshold $p=p^{*}$
[this aspect is reinforced in Fig.~\ref{fig5}(d) 
where we present a phase diagram for $p$ slightly larger than $p=p^{*}$]. 
The fact that the intermediate region  
gets reduced for increasing values of $\sigma$ is expected, since one of
the effects introduced by the parameter $\sigma$ is to destroy several
types of multicritical behavior.  
In Fig.~\ref{fig5}(a), region I is associated with phase diagrams that
present a single 
critical end point, like the one of Fig.~\ref{fig5}(b), whereas region II
is associated with phase diagrams presenting a single ordered critical
point, as shown in Fig.~\ref{fig5}(d); one may go from region I to region
II through the sequence of phase diagrams presented in  
Figs.~\ref{fig5}(b)--(d).  
The ranges of $p$ values associated with these two regions diminish
for increasing values 
of $\sigma/J$, as shown in Fig.~\ref{fig5}(a); 
the computed points along the full lines (empty squares)
represent upper limits (in $\sigma/J$) for regions I and II, in which cases the
corresponding critical points appear at zero temperature. As typical
examples, one has the ranges for the appearence of these points, 
$\sigma/J \leq 0.08 \ (p=0.1)$ and 
$\sigma/J \leq 0.16 \ (p=0.2)$, in region I, whereas 
$\sigma/J \leq 0.2 \ (p=0.5)$ and 
$\sigma/J \leq 0.127 \ (p \approx 0.6667)$, in region II. 
Therefore, for a given pair of parameters $(p,\sigma/J)$ one may predict
the qualitative form of its corresponding phase diagram by using 
Fig.~\ref{fig5}(a). 
Let us first consider one value of $p$ within the range 
$0 < p \lesssim 0.24$, where one may have both a critical end point and a
tricritical point, e.g., the case $p=0.2$, discussed previously. Then, for
sufficiently low values of $\sigma/J$ one has a phase diagram like the one
shown in Fig.~\ref{fig1}(b), characterized by a tricritical point, a
first-order line at low temperatures and a critical end point, associated
with a phase {\bf F}$\!_2$; by increasing $\sigma/J$, one reaches the
upper limit of region I [represented by the continuous line in
Fig.~\ref{fig5}(a)] that corresponds to a collapse of the 
critical end point with the zero-temperature axis, like shown in 
Fig.~\ref{fig1}(c). Increasing $\sigma/J$ further, in such a way that the
probability distribution for the fields still presents a minimum at $H=0$,
one has a tricritical point at finite temperatures [with a typical phase
diagram exhibited in Fig.~\ref{fig1}(d)], and after that, 
one gets phase diagrams like those shown in Figs.~\ref{fig1}(e)
and (f). This sequence of phase diagrams occurs for all values of $p$
within this range.
Now, if one considers $0.27 \lesssim p < p^{*}$ in Fig.~\ref{fig5}(a), 
by increasing $\sigma$ gradually one gets
first a typical phase diagram characterized by an ordered critical point,
like the one shown in Fig.~\ref{fig2}(b), and then, one
may go through the intermediate region between the two dashed lines, 
with a phase diagram as presented in Fig.~\ref{fig5}(c), and
afterwards, one follows a similar sequence of phase diagrams like those 
shown in Figs.~\ref{fig1}(b)--(f).
The last two steps of the previous sequence of phase diagrams apply if one
increases $\sigma/J$ in the range $0.24<p<0.27$.  
By varying $\sigma/J$ for $p$ within the range 
$p^{*} < p \lesssim 0.93$ one gets phase diagrams that follow 
the sequence shown in Figs.~\ref{fig3}. Finally, for 
$0.93 \lesssim p \leq 1.0$ one has a phase diagram typical of the Gaussian
RFIM, i.e., a continuous phase transition separating the paramagnetic and
ferromagnetic phases. 

\subsection{Zero-Temperature Critical Frontiers} 

Some analytical results may be obtained from the investigation of the free
energy and magnetization at zero temperature, given by Eqs.~(\ref{16}) and 
(\ref{17}), respectively. Herein, we shall restrict ourselves to $\sigma>0$ in  
Eq.~(\ref{2}); the case $\sigma=0$, i.e., a trimodal PDF requires a
separate analysis and the results are already known \cite{mattis,kaufman}. 

For that purpose, one applies the same procedure described above for finite
temperatures, starting with the expansion of Eq.~(\ref{17}) in 
powers of $m$, 

\begin{equation} \label{18}
m=a_{1}m+a_{3}m^{3}+a_{5}m^{5}+a_{7}m^{7}+O(m^{9})~, 
\end{equation}

\vskip \baselineskip
\noindent
where

\begin{eqnarray} \label{19}
a_{1} &=&
\sqrt{\frac{2}{\pi}}\left(\frac{J}{\sigma}\right)\left\{p
+(1-p)\exp\left(-\frac{H_{0}^{2}}{2\sigma^{2}}\right)\right\}~, \\  \label{20}
a_{3} &=& \frac{1}{6}\sqrt{\frac{2}{\pi}}\left(\frac{J}{\sigma}\right)^{3}
\left\{(1-p)\left[\left(\frac{H_{0}}{\sigma}\right)^{2}-1\right]
\exp\left(-\frac{H_{0}^{2}}{2\sigma^{2}}\right)-p\right\}~, \\ \label{21}
a_{5} &=& \frac{1}{120}\sqrt{\frac{2}{\pi}}\left(\frac{J}{\sigma}\right)^{5}
\left\{(1-p)\left[\left(\frac{H_{0}}{\sigma}\right)^{4} -
6\left(\frac{H_{0}}{\sigma}\right)^{2}+3\right]
\exp\left(-\frac{H_{0}^{2}}{2\sigma^{2}}\right)+3p\right\} \\ \nonumber
a_{7} &=& \frac{1}{5040}\sqrt{\frac{2}{\pi}}\left(\frac{J}{\sigma}\right)^{7}
\Bigg\{(1-p)\left[\left(\frac{H_{0}}{\sigma}\right)^{6} -
15\left(\frac{H_{0}}{\sigma}\right)^{4}+45\left(\frac{H_{0}}{\sigma}
\right)^{2}-15\right]  \\ \label{22}
& &\times\exp\left(-\frac{H_{0}^{2}}{2\sigma^{2}}\right) - 15p\Bigg\}~.
\end{eqnarray}

\vskip \baselineskip
\noindent
A continuous critical frontier occurs at zero temperature according to the
conditions, $a_{1}=1$ and $a_{3}<0$, leading to a relation 
involving $H_{0}/J$, $\sigma/J$ and $p$, 

\begin{equation} \label{23}
\frac{\sigma}{J}=\sqrt{\frac{2}{\pi}}\left\{p+(1-p)\exp\left[-\frac{1}{2}
\left(\frac{H_{0}}{J}\right)^{2}
\left(\frac{J}{\sigma}\right)^{2}\right]\right\}~.
\end{equation}

\vskip \baselineskip
\noindent
One notices that, for $(H_{0}/J)=0$ one has 
$(\sigma/J) = \sqrt{2/\pi} \cong 0.7979 \ (\forall p)$, yielding a
continuous critical frontier for small values of $H_{0}/J$.  
For $a_{3}>0$, one gets a first-order critical
frontier at zero-temperature, which is usually associated with higher-order
critical points at finite temperatures. A tricritical point appears  
at zero-temperature, provided that $a_{1}=1$, $a_{3}=0$, and $a_{5}<0$,
in such a way that,  

\begin{equation} \label{24}
\frac{H_{0}}{\sigma}=\left[1-\sqrt{\frac{2}{\pi}}\left(\frac{J}{\sigma}
\right)p\right]^{-1/2}~. 
\end{equation}

\vskip \baselineskip
Let us now analyze the coefficient $a_{5}$ under the conditions 
$a_{1}=1$, $a_{3}=0$. Substituting
Eqs~(\ref{23}) and (\ref{24}) in Eq.~(\ref{21}), one gets 

\begin{equation}\label{25}
a_{5}=\frac{1}{120}\left(\frac{J}{\sigma}\right)^{4}
\left\{\left[1-\sqrt{\frac{2}{\pi}}\left(\frac{J}{\sigma}
\right)p\right]^{-1}-3\right\}~,
\end{equation}

\vskip \baselineskip
\noindent
leading to $a_{5}<0$ for
$\left[1-\frac{2}{\pi}\left(\frac{J}{\sigma}\right)p\right]^{-1}$ $<$ $3$.
One should notice that for $p=0$, i.e., for a double-Gaussian PDF
\cite{nuno08a}, the coefficient $a_{5}$ is always negative; however, in the
present case one may have $a_{5}=0$, in such a way that one has the
possibility of a fourth-order critical point. This zero-temperature point
is unique for the distribution  
of  Eq.~(\ref{2}) and it occurs for a set of parameters
$(p^{*},H_{0}^{*}/J,\sigma^{*}/J)$, to be determined below. Considering 
$a_{5}=0$, 

\begin{equation}\label{26}
\frac{\sigma}{J}=\frac{3}{2}\sqrt{\frac{2}{\pi}} \ p~,
\end{equation}

\vskip \baselineskip
\noindent
and using this result in Eq.~(\ref{24}), one gets

\begin{equation}\label{27}
\frac{H_{0}}{\sigma}=\sqrt{3}~. 
\end{equation}

\vskip \baselineskip
\noindent
Taking $a_{3}=0$ and using Eqs.~(\ref{26}) and (\ref{27}) leads to

\begin{equation}\label{28}
p^{*}=2~(2+e^{3/2})^{-1}\cong 0.308561~, 
\end{equation}

\vskip \baselineskip
\noindent
which may be substituted in Eqs.~(\ref{26}), (\ref{27}) to yield, respectively,  

\begin{equation}\label{29}
\frac{\sigma^{*}}{J}=\frac{6}{\sqrt{2\pi}}~(2+e^{3/2})^{-1}\cong 0.369294~; 
\quad
\frac{H^{*}_{0}}{J}=3\sqrt{\frac{6}{\pi}}~(2+e^{3/2})^{-1}\cong 0.639637~.  
\end{equation}

\begin{figure}[t]
\begin{center}
\includegraphics[width=0.5\textwidth,angle=0]{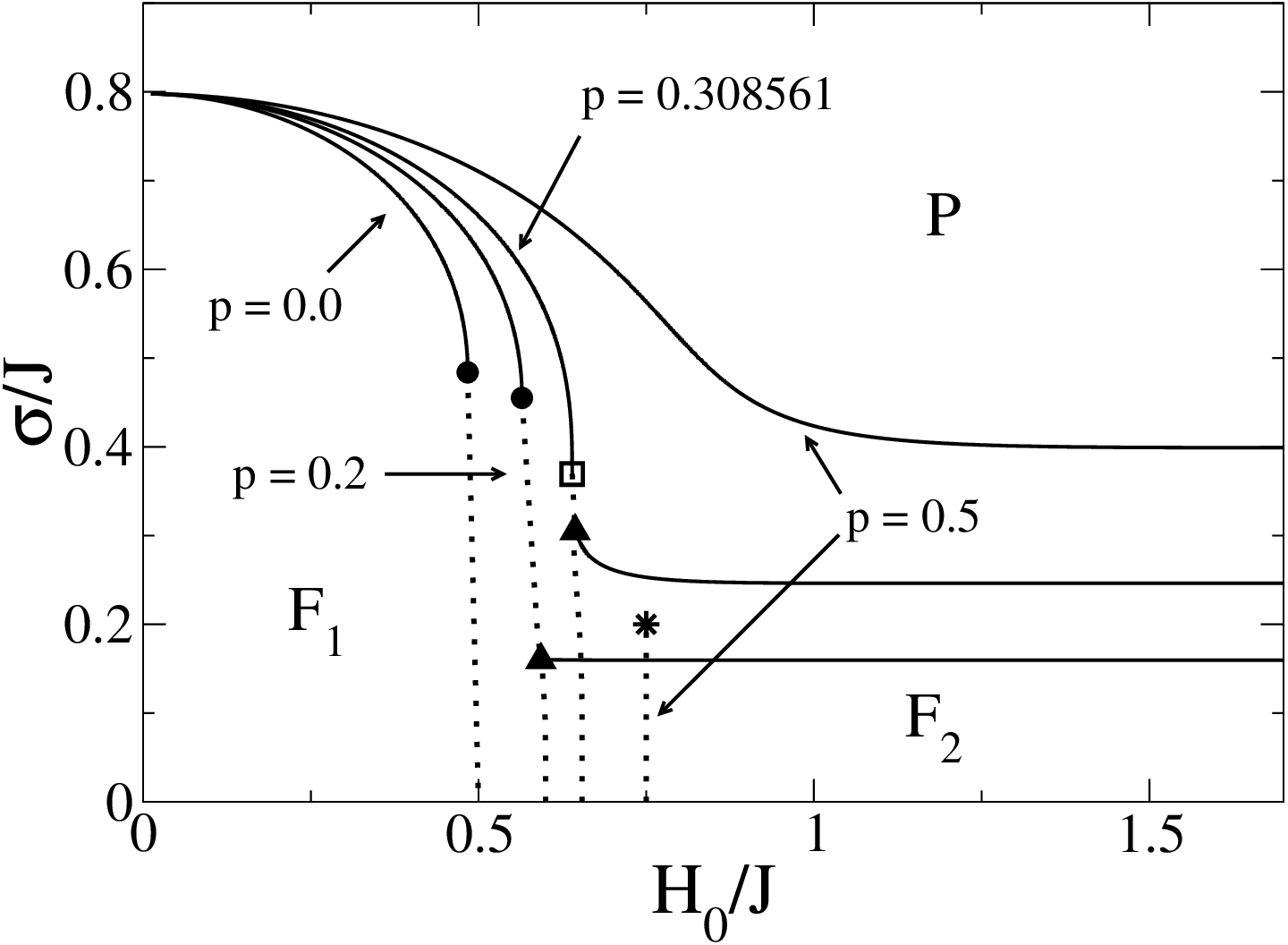}
\end{center}
\vspace{1cm}
\protect\caption{Zero-temperature phase diagrams of the
infinite-range-interaction ferromagnet in the presence of a 
triple Gaussian random field, for typical values of $p$. 
The critical frontiers separate the paramagnetic ({\bf P})
and ferromagnetic ({\bf F}$\!_1$ and {\bf F}$\!_2$) phases.
Critical frontiers and critical points are as described in the text.} 
\label{fig6}
\end{figure}

\vskip \baselineskip
\noindent
Therefore, for the set of parameters of Eqs.~(\ref{28}) and (\ref{29}) one
has a fourth-order critical point at zero temperature, as shown previously
in Fig.~\ref{fig2}(d), and also exhibited in the zero-temperature phase
diagram of Fig.~\ref{fig6}.
In the present problem, the zero-temperature fourth-order critical point 
may be interpreted
as the threshold for the existence of tricritical points in both finite and
zero temperatures. For $p>p^{*}$, there is no tricritical point for
arbitrary values  
of $\sigma/J$ and $H_{0}/J$, although an ordered critical point may still
occur for finite temperatures [cf. Fig.~\ref{fig3}(b)], as well as at zero
temperature    
[see Fig.~\ref{fig6}]. This threshold is associated with a very 
flat PDF, as shown in Fig.~\ref{fig2}(a), which is in agreement with the
conditions for the existence of tricritical points   
\cite{aharony78,andelman,galambirman}. 

Zero-temperature phase diagrams of the model are presented in
Fig.~\ref{fig6} for several values of $p$. One notices that besides
tricritical, and the above-mentioned fourth-order point, critical end
points and ordered critical 
points also appear at $T=0$. These two later points were determined by an
analysis of the 
zero-temperature free energy of Eq.~(\ref{16}), similarly to what was done for
finite temperatures. 
Comparing the phase diagrams of Fig.~\ref{fig6} with those for finite
temperatures, one notices that the width $\sigma$ produces disorder,
playing a role at $T=0$ that is similar to the temperature; as examples,
one sees that the phase diagrams for $p=0$, $p=0.2$, and $p=0.5$, in
Fig.~\ref{fig6}, resemble those shown in  
Figs.~\ref{fig1}(d), \ref{fig1}(b), and \ref{fig3}(b), respectively, if one
considers the correspondence $\sigma \leftrightarrow T$.  

\section{CONCLUSIONS}

We have investigated a random-field Ising model that consists of an 
infinite-range-interaction Ising ferromagnet in the presence of a random
magnetic field following a triple-Gaussian probability distribution. 
Such a distribution, which  is defined as a superposition of
three Gaussian distributions with the same width $\sigma$, 
centered at $H=0$ and $H=\pm H_{0}$, with
probabilities $p$ and $(1-p)/2$, respectively, is very general and recovers 
as limiting cases, the trimodal, bimodal, and Gaussian probability 
distributions. 
We have shown that this model is capable of exhibiting a rich variety of
multicritical phenomena, essentially all types of critical phenomena
found in previous RFIM investigations (as far as we know). We have obtained 
several phase diagrams, for finite temperatures,  
by varying the parameters of the corresponding probability distribution, 
e.g., $p$ and $\sigma$, with variations in $p$ ($\sigma$ fixed) leading to
qualitatively distinct multicritical phenomena, whereas increasing $\sigma$
($p$ fixed) yields a smearing of an specific type of critical behavior.  

The random-field Ising model defined
in terms of a trimodal probability distribution \cite{mattis,kaufman} 
represents, 
to our knowledge, the previously investigated model that exhibits a variety of
multicritical phenomena comparable to the one discussed above. Since the
former represents a particular case ($\sigma=0$) of the present model, 
the parameter $\sigma$ may be considered as an 
additional parameter used herein, when compared with the model of 
Refs.~\cite{mattis,kaufman}. It is important to stress the relevance of
this additional parameter for the richness of the phase diagrams, as well
as for possible physical applications. As one illustration of the phase
diagrams, one may
mention those obtained through the zero-temperature analysis of the
present model, where one gets a large variety of critical frontiers, 
exhibiting all types of critical points that occur at finite temperatures;
this should be contrasted with the much simpler zero-temperature phase
diagram of the trimodal random-field Ising model (where only an ordered
critical point is possible).  In particular,  
some of the zero-temperature phase diagrams presented herein resemble 
those for finite temperatures, if
one considers the correspondence $\sigma \leftrightarrow T$.   
Therefore, $\sigma$ may be identified as a parameter directly related to the
disorder in a real system, in such a way that an increase in  
$\sigma$ in the present model should play a similar role to an increase in
the dilution for a diluted antiferromagnet \cite{nuno08a}. 
It is important to remind that in the identifications of the RFIM with
diluted antiferromagnets 
\cite{fishmanaharony,pozenwong,cardy}, the effective random field at a 
given site is expressed always in terms of quantities that vary in both
sign and magnitude, and so the present distribution for the random fields is more appropriate for this purpose. As a direct consequence of this model,   
the smearing of an specific type of critical behavior, due to an
increase in the randomness, i.e., an increase in $\sigma$, or even a
destruction of a possible multicritical phenomena due to this increase --
which represents a very  
common feature in real systems -- is essentially reproduced by the present
model.    
We have shown that the main effects in the phase diagrams produced by an
increase in $\sigma$ are:  
(i) For those phase diagrams presenting two distinct ferromagnetic phases,
it reduces the ferromagnetic phase with higher magnetization and destroys
the one with lower magnetization;  
(ii) All critical points (tricritical, fourth-order, ordered, and critical
end points) are pushed towards lower temperatures, leading to a destruction
of the first-order critical frontiers. 

We have argued that the random-field Ising model defined in terms of a 
triple-Gaussian probability 
distribution is appropriate for a physical description of some diluted
antiferromagnets that in the presence of a uniform external field may
exhibit multicritical phenomena. As an example  
of a candidate for a physical realization of this model, one has the compound 
${\rm Fe_{x}Mg_{1-x}Cl_{2}}$, which has been modeled through an Ising
ferromagnet under 
random fields whose distribution appears to be well-represented by a
superposition of two parts, namely, a trimodal and a 
continuous contribution~\cite{pozenwong}.
This compound presents a first-order line that disappears due to an
increase in the randomness; such an effect has been described recently in
terms of a simpler model \cite{nuno08a}.  
However, it is possible that the diluted antiferromagnet 
${\rm Fe_{x}Mg_{1-x}Cl_{2}}$, or another similar compound, may present an
even more complicated critical or multicritical behavior; by adjusting
properly the parameters of the random-field distribution, the model
presented herein should be able to cope with such phenomena. 

\vskip 2\baselineskip

{\large\bf Acknowledgments}

\vskip \baselineskip
\noindent
The partial financial supports from CNPq and Pronex/MCT/FAPERJ (Brazilian
agencies) are acknowledged. 

\vskip 2\baselineskip

\end{document}